\documentclass[paper,notoc,nohyper]{JHEP3}
\usepackage{epsfig}
\usepackage{amsbsy} 
\def\S{S_{\epsilon}}
\def\C{{\cal C}}

\def\M{{\cal m}}

\def\Bab{\,{\rm Bub}(\sab)}
\def\Bac{\,{\rm Bub}(\sac)}
\def\Bbc{\,{\rm Bub}(\sbc)}

\def\xac{\left[\Bac-\Bab\right]}
\def\xbc{\left[\Bbc-\Bab\right]}

\def\Boxabac{\,{\rm Box}^6(\sab,\sac)}

\def\Boxabbc{\,{\rm Box}^6(\sab,\sbc)}
\def\dd{d}

\def\dmc{(\dd-3)}
\def\dmd{(\dd-4)}
\def\dme{(\dd-5)}
\def\dmf{(\dd-6)}
\def\dmg{(\dd-7)}

\def\pa{{\cal D}_1^\dagger}
\def\pb{{\cal D}_2^\dagger}
\def\pc{{\cal D}_3^\dagger}
\def\pd{{\cal D}_4^\dagger}
\def\pe{{\cal D}_5^\dagger}
\def\pf{{\cal D}_6^\dagger}

\def\sab{s_{12}}
\def\sac{s_{13}}
\def\sbc{s_{23}}

\def\e{\epsilon}

\def\lnx{X}
\def\lny{Y}
\def\Ls{S}

\def\Libx{{\rm Li}_2(x)}
\def\Liby{{\rm Li}_2(y)}
\def\Licx{{\rm Li}_3(x)}
\def\Licy{{\rm Li}_3(y)}
\def\Lidx{{\rm Li}_4(x)}
\def\Lidy{{\rm Li}_4(y)}
\def\Lidz{{\rm Li}_4\Biggl(\frac{x-1}{x}\Biggr)}

\def\tou{\frac{t}{u}}
\def\one{}

\def\tos{\frac{t}{s}}

\def\CA{C_A}
\def\CF{C_F}

\def\TR{T_R}
\def\NF{N_F}

\def\Mtwid{|M\rangle}
\def\Mt{|\overline {\cal M}\rangle}
\def\M{|{\cal M}\rangle}

\def\C{{\cal C}}
\renewcommand\O[1]{{\cal O}\left(#1\right)}
\def\as{\ensuremath{\alpha_{s}}}
\def\a0{\alpha_0}

\def\Re{\mathop{\rm Re}}

\def\beq{\begin{equation}}
\def\eeq{\end{equation}}

\def\beqn{\begin{eqnarray}}
\def\eeqn{\end{eqnarray}}

\def\lq{\left[}
\def\rq{\right]}

\def\({\left(}
\def\){\right)}

\def\ket#1{|{#1}\rangle}
\def\bra#1{\langle{#1}|}
\def\braket#1#2{\langle #1 |#2 \rangle}
\def\cm{{\cal M}}

\def\MSbar{$\overline{{\rm MS}}$}

\def\bom#1{{\mbox{\boldmath $#1$}}}

\def\fs{\(-\frac{\mu^2}{\sab}\)^\ep }
\def\ft{\(-\frac{\mu^2}{\sbc}\)^\ep }
\def\fu{\(-\frac{\mu^2}{\sac}\)^\ep }
\def\fsd{\(-\frac{\mu^2}{\sab}\)^{2\ep} }
\def\ftd{\(-\frac{\mu^2}{\sbc}\)^{2\ep} }
\def\fud{\(-\frac{\mu^2}{\sac}\)^{2\ep} }
\def\ep{\epsilon}

\def\Ls{L_s}

\def\Lx{L_x}
\def\Ly{L_y}
\def\Lx{X}
\def\Ly{Y}
\def\Ls{S}

\def\Libx{{\rm Li}_2(x)}
\def\Liby{{\rm Li}_2(y)}
\def\Licx{{\rm Li}_3(x)}
\def\Licy{{\rm Li}_3(y)}

\def\Lidx{{\rm Li}_4(x)}
\def\Lidy{{\rm Li}_4(y)}
\def\Lidz{{\rm Li}_4(z)}

\def\TTOSS{\frac{t^2}{s^2}}
\def\TTOUU{\frac{t^2}{u^2}}

\def\UUOTT{\frac{u^2}{t^2}}

\def\tos{\frac{t}{s}}
\def\tou{\frac{t}{u}}

\def\uot{\frac{u}{t}}
\def\xxx{\nonumber \\ &&}

\title{\boldmath 
Two-loop QCD helicity amplitudes for massless quark-quark scattering
}
\author{
E.~W.~N.~Glover$^a$\\
$^a$Department of Physics, 
University of Durham, 
Durham DH1 3LE, 
England\\ 
E-mail:  \email{E.W.N.Glover@durham.ac.uk}}
\abstract{
We present the two-loop helicity amplitudes for the scattering of massless
quarks in QCD.
We use projector techniques to compute the coefficients of the general
four-quark amplitude at up to two-loops in conventional
dimensional regularisation and use these coefficients to derive the
helicity amplitudes in the 't Hooft-Veltman scheme. The structure of the
infrared divergences agrees with that predicted by Catani while
expressions for the finite remainders for $q \bar q \to Q\bar Q$ 
and the processes related by crossing symmetry
 are given in
terms of logarithms and  polylogarithms that are real in the physical
region.  We have checked that the interference of tree and two-loop helicity
amplitudes, summed over helicities and colours, reproduces 
the previous results for the finite remainders for 
interference of tree and two-loop amplitudes
given in Ref.~[1,2].  
}

\preprint{{DCTP/04/04}, {IPPP/04/02}, {hep-ph/0401119}}

\begin{document}
%%%%%to make all the references appear 
%%%%%eventhough they are not in use
%%%%%\nocite{*}
\nocite{qqunlike,qqlike}

\section{Introduction}
\label{sec:intro}

In the past decade, QCD has become a quantitative science and comparisons of
high quality experimental data from high energy collider experiments  with QCD
predictions  at next-to-leading order (NLO) in $\as$ have become the {\it de
facto} standard. In such studies, the parton level predictions serve  to
provide an estimate of the uncalculated theoretical uncertainties via the
variation in the renormalisation (and factorisation) scale. However, there are
many well established reasons why extending perturbative calculations to
next-to-next-to-leading order  (NNLO) is vital in reducing the theoretical
uncertainty for the dominant $2 \to 1$   and $2 \to 2$ processes such as
Drell-Yan or jet production.   Most of the anticipated improvements are either
due  (a) to including the next term in the perturbation series or  (b) to
including the effect of more partons in the final state. Here, we list six of
the improvements expected for NNLO predictions at hadron colliders and  refer
the reader to Ref.~\cite{nigelRCLL} for a more detailed discussion;
\begin{enumerate}
\item (a) Reduction of the renormalisation scale uncertainty;
\item (a) Reduction of the factorisation scale uncertainty;
\item (a) Reduced power corrections;
\item (b) Better matching between parton-level and hadron level jet algorithms;
\item (b) Better description of the transverse momentum of the initial state;
\item (b) Extended kinematic coverage due to enlarged phase space.
\end{enumerate}
All of these will be underpinned by a better knowledge of the parton
distributions  using global fits to NNLO observables from DIS, Drell-Yan and
jet production and by a better determination of $\as$ obtained from 3~jet event
shape data at electron-positron colliders.  Taken together, one expects that
the theoretical uncertainty on a given observable will drop from 
${\cal O}(20\%)
$ at NLO to  
$<{\cal O}(10\%) 
$ at NNLO.

In this paper, we focus on jet production in hadron-hadron collisions.
The full NNLO prediction requires a knowledge of the
two-loop $2\to 2$ partonic matrix elements, as well as the 
one-loop $2\to 3$ and tree-level $2\to 4$ amplitudes and NNLO parton distributions.
Much work has been invested in the three-loop splitting
functions~\cite{moms1,moms2,moms3,Gra1,NV1,NV2,NVplb,VMVdis,carola,oneloopsplit,Kosower:evolution} needed 
to  evolve the parton distribution functions at this order.  The available moments have been
used to provide approximate fits in $x$-space and NNLO parton distributions
 are beginning to become available~\cite{MRST,MRST-NNLO}.  The exact three-loop
 splitting functions have been eagerly anticipated and the unpolarised singlet and non-singlet
 splitting functions are now available~\cite{singlet,nsinglet}. 
On the other hand, the tree-level six-point amplitudes~\cite{6g1,6g2,6g3,6g4,4g2q1,4g2q2,2g4q,6q} and
one-loop five point amplitudes~\cite{5g,3g2q,1g4q} have been known for some time
while the interference of tree and two-loop amplitudes~\cite{qqunlike,qqlike,qqgg,gggg}
and self-interference of the one-loop amplitudes~\cite{qqgg,qqqq1Lsq,gggg1Lsq} are also
available.
Subsequently, two-loop helicity amplitudes have been evaluated for the 
gluon-gluon~\cite{BDKgggg,BFDgggg} and quark-gluon~\cite{BFDqqgg,qqgghel} processes
and have confirmed the earlier "squared" matrix elements.\footnote{Note that helicity amplitudes for the phenomenologically important $gg \to
\gamma\gamma$ process~\cite{BFDggpp}, $e^+e^- \to 3$~jets~\cite{helamp3j} and $\gamma\gamma \to
\gamma\gamma$~\cite{BFDGpppp,BGMBpppp} have also been computed.}
Here, we complete the set of two-loop QCD helicity amplitudes for parton-parton scattering 
by studying processes involving two pairs of massless 
quarks,
$$ q\bar q \to Q \bar Q,$$
and the processes related  by time reversal and crossing symmetry when 
$q$ and $Q$ are either distinct or
identical. 
 
We extract the helicity amplitudes using the same method as that employed in
Refs.~\cite{BGMBpppp,helamp3j,qqgghel}.   First we write down a general tensor that
describes the four quark amplitude.   Because the Dirac algebra is infinite
dimensional for non-integer $\dd$,  we use a basis set of Dirac strings that are
sufficient to describe the amplitude up to two-loops.  At three-loops new structures
will appear.  The coefficients are then extracted using a set of projectors that are
valid at up to two-loops. These projectors turn the Dirac strings into traces (which
are taken in $\dd$-dimensions) and  thereby reduce the problem to evaluating scalar
integrals. The methodology of reducing these scalar integrals to master integrals is
exactly the same as for the calculation of the spin summed two-loop matrix
elements~\cite{qqunlike,qqlike}.   By taking the trace in $\dd$-dimensions and 
keeping all Lorentz indices in $\dd$-dimensions, we ensure that tensor coefficients are evaluated
in conventional dimensional regularisation (CDR).
We then provide the perturbative expansion for the one- and two-loop tensor
coefficients and remove ultraviolet  (UV) divergences at each order in $\as$ by
renormalisation within the \MSbar~scheme. The infrared (IR) divergent structure is
shown to agree with that obtained in Ref.~\cite{catani}. 
Finally, the helicity amplitudes are obtained from the general tensor using standard
helicity techniques and within the t'Hooft-Veltman (HV) scheme where the external
states are four-dimensional and the internal states are kept in
$\dd=4-2\epsilon$-dimensions~\cite{hv1,hv2}.  The finite helicity amplitudes are the
main new results presented in this paper and we give explicit analytic expressions
valid for each helicity configuration and process in terms of logarithms and
polylogarithms that are real in the physical domain.

The paper is organised as follows. In Section~\ref{sec:notation} we introduce
notation and discuss the general tensor structure of the four-quark amplitude at up
to two-loops.     We give projectors to isolate the various coefficients and relate
the individual coefficients to the interference of amplitudes. 
Section~\ref{sec:perturbative}  shows the perturbative expansion for the tensor
coefficients computed within CDR and presents an analysis of their UV and IR
divergent structure. The coefficients are renormalised in the \MSbar~scheme to
remove all UV divergences, while the IR poles are predicted using the results of
Ref.~\cite{catani}.   Expressions for the one-loop tensor coefficients to all orders
in $\epsilon$ appear in Section~\ref{subsec:oneloop} in terms of the one-loop bubble
integral and the one-loop box integral in $\dd=6-2\epsilon$.    The relationship
between the tensor coefficients and the helicity amplitudes  in the HV scheme is
described in Section~\ref{sec:helamp}.   The UV and IR structure of the helicity
amplitudes is exactly the same as for the tensor coefficients, and finite
$\O{\ep^0}$ remainders are obtained by subtracting the  predicted pole structure
from the explicit calculation both at one-loop and two-loop.   The main results of
this paper are the finite remainders of the two-loop helicity amplitudes.   The
expressions are rather lengthy and we present the amplitudes for the four-quark
process in Appendix~\ref{app:twoloop}.  For completeness, the finite one-loop
remainders are collected in Appendix~\ref{app:oneloop}.   Finally, in
Section~\ref{sec:conclusions} we summarize our results.

\newpage
\section{Notation}
\label{sec:notation}

The processes that we wish to consider in detail are,
\begin{eqnarray}
\label{eq:qqQQ}
0 &\to& q \bar q Q\bar Q,\\
\label{eq:qqqq}
0 &\to& q \bar q q \bar q .
\end{eqnarray}
Because the identical quark amplitudes are obtained from the non-identical quark
amplitudes by the exchange of the quark or antiquark momenta, we focus on the
scattering of distinct quarks,
\begin{equation} 
0 \to q(p_1,\lambda_1) + \bar q(p_2,\lambda_2) + Q(p_3,\lambda_3) + \bar Q(p_4,\lambda_4),
\end{equation}
where the quarks $q$ and $Q$ are taken to be massless
with momenta satisfying,
\begin{equation}
0 \to p_1^\mu +p_2^\mu +p_3^\mu +p_4^\mu, \qquad \qquad p_i^2 = 0.
\end{equation}
Physical processes are obtained by crossing particles into the initial state.
The associated Mandelstam variables are given by
\begin{equation}
s_{12} = (p_1+p_2)^2, \qquad s_{23} = (p_2+p_3)^2, \qquad s_{13} = (p_1+p_3)^2, 
\qquad s_{12}+s_{23}+s_{13} = 0.
\end{equation}

We work in conventional dimensional regularisation and renormalise the ultraviolet 
divergences in the \MSbar\ scheme. The bare coupling $\a0$ is related to 
the running coupling $\as \equiv \alpha_s(\mu^2)$  at renormalisation 
scale $\mu$, by
\beq
\label{eq:alpha}
\a0 \mu_0^{2\e} \,  \S = \as \,  \mu^{2\e} \, \lq 1 - \frac{\beta_0}{\ep}  
\, \left(\frac{\as}{2\pi}\right) + \( \frac{\beta_0^2}{\ep^2} - \frac{\beta_1}{2\ep} \)  \, 
\left(\frac{\as}{2\pi}\right)^2
+\O{\as^3} \rq,
\eeq
where
\beq
\S = (4 \pi)^\ep e^{-\ep \gamma},  \quad\quad \gamma=0.5772\ldots=
{\rm Euler\ constant}
\eeq
is the typical phase-space volume factor in $d=4-2\ep$ dimensions
and $\mu_0^2$ is the mass parameter introduced 
in dimensional regularisation~\cite{dreg1,dreg2,hv1,hv2} to maintain a 
dimensionless coupling 
in the bare QCD Lagrangian density.

The first two coefficients of the QCD beta function, 
$\beta_0$ and $\beta_1$, for $N_F$ (massless) quark flavours, are
\beq
\label{betas}
\beta_0 = \frac{11 \CA - 4 T_R \NF}{6} \;\;, \;\; \;\;\;\;
\beta_1 = \frac{17 \CA^2 - 10 \CA T_R \NF - 6 \CF T_R \NF}{6} \;\;,
\eeq
where $N$ is the number of colours and 
\beq
\CF = \frac{N^2-1}{2N}, \qquad \CA = N, \qquad T_R = \frac{1}{2}.
\eeq

\subsection{The general tensor}
\label{subsec:genten}
The most general tensor structure for the amplitude, $\M$, for the scattering of
distinct massless quarks at up to two loops can be written as
\begin{eqnarray}
\label{eq:gentensor}
\M &=&  \sum_{I=1}^{6} A_I(\sab,\sbc) ~{\cal D}_I + \ldots
\end{eqnarray}
where the coefficients $A_I$  
are vectors in colour space and are functions of $\sab$ and $\sbc$ (and implicitly $\sac =
-\sab-\sbc$) where $s_{ij}= (p_i+p_j)^2$ and the six Dirac structures are
\begin{eqnarray}
\label{eq:dirac}
{\cal D}_1 &=&
~\bar u(p_1) \gamma_{\mu_1} u(p_2) ~\bar u(p_3) \gamma_{\mu_1} u(p_4),
\nonumber \\
{\cal D}_2 &=&
 ~\bar u(p_1)\slash \!\!\! p_3 u(p_2) ~\bar u(p_3)\slash \!\!\! p_1 u(p_4),
\nonumber \\
{\cal D}_3 &=&
 ~\bar u(p_1) \gamma_{\mu_1} \gamma_{\mu_2}\gamma_{\mu_3} u(p_2) ~\bar
u(p_3)\gamma_{\mu_1}\gamma_{\mu_2}\gamma_{\mu_3}  u(p_4),
\nonumber \\
{\cal D}_4 &=&
 ~\bar u(p_1) \gamma_{\mu_1} \slash \!\!\! p_3\gamma_{\mu_3} u(p_2) ~\bar
u(p_3)\gamma_{\mu_1}\slash \!\!\! p_1\gamma_{\mu_3}  u(p_4),
\nonumber \\
{\cal D}_5 &=&
 ~\bar u(p_1) \gamma_{\mu_1} \gamma_{\mu_2}\gamma_{\mu_3}\gamma_{\mu_4}\gamma_{\mu_5} u(p_2) ~\bar
u(p_3)\gamma_{\mu_1}\gamma_{\mu_2}\gamma_{\mu_3}\gamma_{\mu_4}\gamma_{\mu_5} 
u(p_4),
\nonumber \\
{\cal D}_6 &=&
 ~\bar u(p_1) \gamma_{\mu_1} \gamma_{\mu_2}\slash \!\!\! p_3\gamma_{\mu_4}\gamma_{\mu_5} u(p_2) ~\bar
u(p_3)\gamma_{\mu_1}\gamma_{\mu_2}\slash \!\!\! p_1\gamma_{\mu_4}\gamma_{\mu_5} 
u(p_4).
\end{eqnarray}
This tensor structure is a priori $\dd$-dimensional since the Lorentz indices 
are $\dd$-dimensional and the dimensionality (and helicity) of the external
states has not yet been specified.  One can in principle relate the strings of
gamma matrices appearing in ${\cal D}_3$ to ${\cal D}_6$ to a standard set
involving  only ${\cal D}_1$ and ${\cal D}_2$ using four-dimensional tricks.  
However, because these are the structures that naturally arise in the parity
conserving interactions of QCD, we choose to use this extended set as a
$\dd$-dimensional basis that is valid at up to two-loops. We note that the
Dirac algebra is infinite dimensional for non-integer $\dd$ and that the basis set
will extend according to the order that $\M$ is computed.    For example, at
tree level, only ${\cal D}_1$ appears, while ${\cal D}_2$, ${\cal D}_3$ and
${\cal D}_4$ first appear at one-loop.  ${\cal D}_5$ and ${\cal D}_6$ appear
for the first time at two-loops while at three-loops, we will find terms
(represented by $+\ldots$) with seven gamma matrices sandwiched between the
quark spinors.  These more complicated structures can also be related to the
simpler ones using four-dimensional tricks (which we choose not to do at the
present time).   

When the quarks are identical, the general structure of the amplitude is
modified,
\begin{equation}
\label{eq:Mident}
\Mtwid = \M - \delta_{qQ}\Mt,
\end{equation}
where
\begin{equation}
\Mt = \M (p_2 \leftrightarrow p_4).
\end{equation}
The minus sign is due to the exchange of identical fermions, while the momentum swap
corresponds to exchanging $\sab$ and $\sbc$ in the coefficents $A_I$.
All appropriate colour indices are also exchanged.
In general we will multiply these additional identical fermion terms with a
$\delta_{qQ}$ which is unity when the quarks are identical and zero otherwise.

\subsection{Projectors for the tensor coefficients}
\label{subsec:projectors}

The six coefficients $A_I$  may be easily extracted from a 
Feynman diagram calculation with two distinct quark flavours
using projectors that act on the general tensor of Eq.~(\ref{eq:gentensor}) 
such that
\begin{equation}
\sum_{\rm spins} {\cal P}(A_I) ~\M  = A_I(\sab,\sbc ).
\end{equation}
The explicit forms for the projectors in $\dd$ space-time dimensions are,
\begin{eqnarray}
{\cal P}(A_1) &=&\frac{1}{480\sac\sbc^2\sab^2\dme\dmf\dmg\dmc\dmd}\times \biggl (\\\nonumber
&+&\sac(-9240\sbc^2\dd^3-35040\sac\sbc\dd^2+52160\sac^2\dd-61120\sac^2+61620\sbc^2\dd^2
\\\nonumber &&
+164320\sac\sbc\dd+202496\sbc^2-12720\sac^2\dd^2-182480\sbc^2\dd+960\sac^2\dd^3-259840\sac\sbc
\\\nonumber &&
+525\sbc^2\dd^4+2520\sac\sbc\dd^3)\pa \\\nonumber
&-&10\sac(24\sac^2\dd^2-952\sac\sbc\dd+102\sac\sbc\dd^2-1568\sbc^2+2344\sac\sbc-264\sbc^2\dd^2\\\nonumber &&
+21\sbc^2\dd^3+256\sac^2-176\sac^2\dd+1124\sbc^2\dd)\pc \\\nonumber
&-&15\dmf(35\sbc^2\dd^3-55\sac\sbc\dd^3+1046\sac\sbc\dd^2-1872\sac^2\dd+2432\sac^2-454\sbc^2\dd^2\\\nonumber &&
-6040\sac\sbc\dd-2688\sbc^2+368\sac^2\dd^2+1928\sbc^2\dd-20\sac^2\dd^3+11136\sac\sbc)\pb \\\nonumber
&+&\sac\sbc(-320\sac+15\sbc\dd^2-110\sbc\dd+224\sbc+60\sac\dd)\pe \\\nonumber
&-&5(-102\sbc^2\dd+15\sbc^2\dd^2-1048\sac\sbc+168\sbc^2+88\sac^2\dd-128\sac^2-27\sac\sbc\dd^2
\\\nonumber &&
+326\sac\sbc\dd-12\sac^2\dd^2)\pf \\\nonumber
&+&30(21\sbc^2\dd^3-37\sac\sbc\dd^3+672\sac\sbc\dd^2-1104\sac^2\dd+1360\sac^2-256\sbc^2\dd^2-3868\sac\sbc\dd\\\nonumber &&
-1344\sbc^2+244\sac^2\dd^2+1036\sbc^2\dd-16\sac^2\dd^3+7328\sac\sbc)\pd \biggr),\\\nonumber
{\cal P}(A_2) &=&\frac{1}{32\sac^2\sbc^2\sab^2\dme\dmg\dmc\dmd}\times \biggl (\\\nonumber
&-&\sac(35\sbc^2\dd^3-55\sac\sbc\dd^3+1046\sac\sbc\dd^2-1872\sac^2\dd+2432\sac^2-454\sbc^2\dd^2
\\\nonumber &&
-6040\sac\sbc\dd-2688\sbc^2+368\sac^2\dd^2+1928\sbc^2\dd-20\sac^2\dd^3+11136\sac\sbc)\pa \\\nonumber
&+&2\sac(-2\sac^2\dd^2-9\sac\sbc\dd^2+142\sac\sbc\dd-448\sac\sbc+7\sbc^2\dd^2+136\sbc^2-48\sac^2\\\nonumber &&
+28\sac^2\dd-62\sbc^2\dd)\pc \\\nonumber
&+&(-340\sac^2\dd^3+11008\sac^2-740\sac\sbc\dd^3+44032\sac\sbc-260\sbc^2\dd^3-4144\sbc^2\dd+3712\sbc^2\\\nonumber &&
+15\sac^2\dd^4+2852\sac^2\dd^2-28864\sac\sbc\dd+1604\sbc^2\dd^2+6944\sac\sbc\dd^2-9968\sac^2\dd\\\nonumber &&
+30\sac\sbc\dd^4+15\sbc^2\dd^4)\pb \\\nonumber
&-&\sac\sbc(12\sac+\sbc\dd-4\sbc-\sac\dd)\pe \\\nonumber
&+&(-6\sbc^2\dd+24\sac^2+2\sac\sbc\dd^2-40\sac\sbc\dd-14\sac^2\dd+\sac^2\dd^2+8\sbc^2+\sbc^2\dd^2+192\sac\sbc)\pf \\\nonumber
&-&2(5\sac^2\dd^3+5\sbc^2\dd^3+10\sac\sbc\dd^3-240\sac\sbc\dd^2-100\sac^2\dd^2-56\sbc^2\dd^2+580\sac^2\dd\\\nonumber &&
+1832\sac\sbc\dd+196\sbc^2\dd-208\sbc^2-800\sac^2-4224\sac\sbc)\pd \biggr),\\\nonumber
{\cal P}(A_3) &=&\frac{1}{48\sac\sbc^2\sab^2\dme\dmf\dmg\dmc\dmd}\times \biggl (\\\nonumber
&-&\sac(24\sac^2\dd^2-952\sac\sbc\dd+102\sac\sbc\dd^2-1568\sbc^2+2344\sac\sbc-264\sbc^2\dd^2\\\nonumber &&
+21\sbc^2\dd^3+256\sac^2-176\sac^2\dd+1124\sbc^2\dd)\pa \\\nonumber
&+&2\sac(20\sac\sbc\dd+122\sbc^2+4\sac^2\dd-8\sac^2+6\sbc^2\dd^2-53\sbc^2\dd-80\sac\sbc)\pc \\\nonumber
&+&3\dmf(-2\sac^2\dd^2-9\sac\sbc\dd^2+142\sac\sbc\dd-448\sac\sbc+7\sbc^2\dd^2+136\sbc^2\\\nonumber &&
-48\sac^2+28\sac^2\dd-62\sbc^2\dd)\pb \\\nonumber
&-&\sac\sbc(\sbc\dd-4\sbc+2\sac)\pe \\\nonumber
&+&(-20\sbc^2-7\sac\sbc\dd+5\sbc^2\dd+4\sac^2-2\sac^2\dd+44\sac\sbc)\pf \\\nonumber
&-&6(-2\sac^2\dd^2-8\sac\sbc\dd^2+105\sac\sbc\dd-298\sac\sbc+6\sbc^2\dd^2+122\sbc^2\\\nonumber &&
-32\sac^2+20\sac^2\dd-53\sbc^2\dd)\pd \\\nonumber
{\cal P}(A_4) &=&\frac{1}{16\sac^2\sbc^2\sab^2\dme\dmf\dmg\dmc\dmd}\times \biggl (\\\nonumber
&+&\sac(21\sbc^2\dd^3-37\sac\sbc\dd^3+672\sac\sbc\dd^2-1104\sac^2\dd+1360\sac^2-256\sbc^2\dd^2\\\nonumber &&
-3868\sac\sbc\dd-1344\sbc^2+244\sac^2\dd^2+1036\sbc^2\dd-16\sac^2\dd^3+7328\sac\sbc)\pa \\\nonumber
&-&2\sac(-2\sac^2\dd^2-8\sac\sbc\dd^2+105\sac\sbc\dd-298\sac\sbc+6\sbc^2\dd^2+122\sbc^2-32\sac^2\\\nonumber &&
+20\sac^2\dd-53\sbc^2\dd)\pc \\\nonumber
&-&\dmf(5\sac^2\dd^3+5\sbc^2\dd^3+10\sac\sbc\dd^3-240\sac\sbc\dd^2-100\sac^2\dd^2-56\sbc^2\dd^2+580\sac^2\dd\\\nonumber &&
+1832\sac\sbc\dd+196\sbc^2\dd-208\sbc^2-800\sac^2-4224\sac\sbc)\pb \\\nonumber
&+&\sac\sbc(-\sac\dd-4\sbc+\sbc\dd+8\sac)\pe \\\nonumber
&-&(16\sac^2+144\sac\sbc-36\sac\sbc\dd+2\sac\sbc\dd^2-10\sac^2\dd-6\sbc^2\dd+8\sbc^2+\sbc^2\dd^2+\sac^2\dd^2)\pf \\\nonumber
&+&2(-186\sac\sbc\dd^2+4\sac^2\dd^3+356\sac^2\dd-69\sac^2\dd^2-2976\sac\sbc-468\sac^2-180\sbc^2\\\nonumber &&
+8\sac\sbc\dd^3-45\sbc^2\dd^2+1348\sac\sbc\dd+164\sbc^2\dd+4\sbc^2\dd^3)\pd \biggr),\\\nonumber
{\cal P}(A_5) &=&\frac{1}{480\sac\sbc\sab^2\dme\dmf\dmg\dmc\dmd}\times \biggl (\\\nonumber
&+&\sac(-320\sac+15\sbc\dd^2-110\sbc\dd+224\sbc+60\sac\dd)\pa \\\nonumber
&-&10\sac(\sbc\dd-4\sbc+2\sac)\pc \\\nonumber
&-&15\dmf(12\sac+\sbc\dd-4\sbc-\sac\dd)\pb \\\nonumber
&+&\sac\sbc\pe \\\nonumber
&-&5(-\sac+\sbc)\pf \\\nonumber
&+&30(-\sac\dd-4\sbc+\sbc\dd+8\sac)\pd \biggr),\\\nonumber
{\cal P}(A_6) &=&\frac{1}{96\sac^2\sbc^2\sab^2\dme\dmf\dmg\dmc\dmd}\times \biggl (\\\nonumber
&-&\sac(-102\sbc^2\dd+15\sbc^2\dd^2-1048\sac\sbc+168\sbc^2+88\sac^2\dd-128\sac^2-27\sac\sbc\dd^2\\\nonumber &&
+326\sac\sbc\dd-12\sac^2\dd^2)\pa \\\nonumber
&+&2\sac(-20\sbc^2-7\sac\sbc\dd+5\sbc^2\dd+4\sac^2-2\sac^2\dd+44\sac\sbc)\pc \\\nonumber
&+&3\dmf(-6\sbc^2\dd+24\sac^2+2\sac\sbc\dd^2-40\sac\sbc\dd-14\sac^2\dd+\sac^2\dd^2+8\sbc^2\\\nonumber &&
+\sbc^2\dd^2+192\sac\sbc)\pb \\\nonumber
&-&\sac\sbc(-\sac+\sbc)\pe \\\nonumber
&+&(-2\sac^2+\sac^2\dd+\sbc^2\dd-24\sac\sbc-2\sbc^2+2\sac\sbc\dd)\pf \\\nonumber
&-&6(16\sac^2+144\sac\sbc-36\sac\sbc\dd+2\sac\sbc\dd^2-10\sac^2\dd-6\sbc^2\dd+8\sbc^2+\sbc^2\dd^2+\sac^2\dd^2)\pd 
\biggr),
\end{eqnarray} 
where ${\cal D}_i^\dagger$ is the complex conjugate of the Dirac structures defined in Eq.~(\ref{eq:dirac}).

\subsection{Reconstruction of the matrix elements}

Starting from the tensor coefficients, the interference (or square) of 
the amplitude of Eq.~(\ref{eq:Mident}) can be easily extracted and we
find that,  
\begin{eqnarray}
\label{eq:square}
\sum_{\rm spins} \langle M | M \rangle   &=& 
2 \sac^2 \left| B_1(\sab,\sbc)+\delta_{qQ} B_1(\sbc,\sab) \right|^2 
\nonumber
\\
&+& 2\sbc^2 
\left|B_2(\sab,\sbc)  \right|^2 + 
2\delta_{qQ} \sab^2 \left| B_2(\sbc,\sab)\right|^2 + {\cal O}(\epsilon)
\end{eqnarray}
where the combinations of tensor coefficients $B_1$ and $B_2$ are
\begin{eqnarray}
\label{eq:b1andb2}
B_1(\sab,\sbc) &=&2A_1(\sab,\sbc)-\sbc A_2(\sab,\sbc)+32 A_3(\sab,\sbc)+4\sac A_4(\sab,\sbc)\nonumber \\
&&+512 A_5(\sab,\sbc) 
+ 64\sac A_6(\sab,\sbc),\nonumber \\
B_2(\sab,\sbc) &=& 2A_1(\sab,\sbc)+\sac A_2(\sab,\sbc)+8 A_3(\sab,\sbc)  \nonumber \\
&&+32 A_5(\sab,\sbc)+ 16\sac A_6(\sab,\sbc).
\end{eqnarray}
As mentioned earlier, $\delta_{qQ}=1$ for identical quarks and is zero otherwise. Note
that the tensor coefficients are in general matrices in colour space and the colour sum
has still to be performed in evaluating, for example,   $A_{1}A_{1}^\dagger$.  
The functions $B_i(\sbc,\sab)$ are obtained from $B_i(\sab,\sbc)$ by exchanging the colour indices of the antiquarks as
well as $p_2$ and $p_4$.
Additional terms of order $\epsilon$ and $\epsilon^2$ due to the $\dd$-dimensional
nature of the Lorentz indices present in the six Dirac structures are straightforwardly
generated.  The resulting formula is, however, rather lengthy and we do not give it
here.

Note that the squared amplitude is conventionally denoted
by~\cite{ES,qqunlike,qqlike}
\begin{equation}
\sum_{\rm spins} \langle  M \Mtwid  =
{\cal A}(\sab,\sbc) + \delta_{qQ}{\cal A}(\sbc,\sab) + \delta_{qQ}{\cal B}(\sab,\sbc),
\end{equation}
and by inspection we see that
\begin{eqnarray}
{\cal A}(\sab,\sbc) &=& 2 \sac^2 \left| B_1(\sab,\sbc)\right|^2 +2\sbc^2 
\left|B_2(\sab,\sbc)  \right|^2+ {\cal O}(\epsilon),\nonumber \\
{\cal B}(\sab,\sbc) &=& 2 \sac^2 \left( B_1(\sab,\sbc)B_1^\dagger(\sbc,\sab)
+ B_1(\sbc,\sab)B_1^\dagger(\sab,\sbc)\right)+ {\cal O}(\epsilon).
\end{eqnarray}

\section{The perturbative expansion of the tensor coefficients}
\label{sec:perturbative}

The amplitude $\M$  defined in Eq.~(\ref{eq:gentensor})
has the perturbative expansion of the form,
\begin{eqnarray}
\label{eq:perturb}
|{\cal M}\rangle &=&  4\pi \,\alpha_s \left[
|{\cal M}^{(0)}  \rangle
+ \left(\frac{\alpha_s}{2\pi}\right)|{\cal M}^{(1)}  \rangle
+ \left(\frac{\alpha_s}{2\pi}\right)^2 |{\cal M}^{(2)} \rangle
+ {\cal O}(\alpha_s^3) \right],
\end{eqnarray}
where $ |{\cal M}^{(n)}\rangle$ denotes the $n$-loop contribution to the amplitude.
Similarly, the coefficients have 
perturbative expansions of the form
\begin{eqnarray}
\label{eq:perturbA}
A_I(\sab,\sbc) &=&  4\pi \,\alpha_s \left[
A_I^{(0)} (\sab,\sbc) 
+ \left(\frac{\alpha_s}{2\pi}\right) A_I^{(1)}(\sab,\sbc)  
+ \left(\frac{\alpha_s}{2\pi}\right)^2 A_I^{(2)} (\sab,\sbc)
+ {\cal O}(\alpha_s^3) \right].
\end{eqnarray}
The $n$-loop perturbative coefficients are vectors in colour space and 
can be further decomposed as,
\begin{eqnarray}
\label{eq:colour}
A_I^{(n)}(\sab,\sbc) &=& 
\sum_{i=1}^2 \;\C_i\;A_I^{(n),[i]}(\sab,\sbc),
\end{eqnarray}
where
\begin{equation}
\C_1 = {\bom \delta_{i_3i_2}}{\bom \delta_{i_1i_4}}, \qquad
\C_2 = {\bom \delta_{i_3i_4}}{\bom \delta_{i_1i_2}}, \qquad
\end{equation}
and indices $i_1,\ldots,i_4$ denote colour labels in the fundamental representation.
In evaluating the squared amplitudes, we will encounter the colour sum 
$\sum_{colours} \C_i \C_j^\dagger$ for $1 \leq i,j \leq 2$ which is given by the
matrix ${\cal CC}_{ij}$ where
\begin{equation}
\label{eq:CC}
{\cal CC}_{ij} = 
\(\begin{array}{cc}
N^2 & N \\
N & N^2
\end{array}
\).
\end{equation}

As in previous work, we use QGRAF~\cite{QGRAF} to generate the Feynman diagrams
and then use MAPLE~\cite{MAPLE} and FORM~\cite{FORM3} to manipulate the Dirac
structures and perform the $\dd$-dimensional traces. When the projectors are
applied to the individual Feynman diagrams, and the traces taken, we are left
with a collection of two-loop scalar integrals with different topologies and
powers of propagators. It is now well known (see for
example~\cite{houchesQCD,thomasRCLL,thomasHAD} and references therein), that
these integrals can be reduced~\cite{laporta} to a handful of {\it master
integrals} using integration-by-parts~\cite{IBP1,IBP2}  and
Lorentz-invariance~\cite{diffeqLI} identities. For the process at hand --
two-loop box graphs with four on-shell legs -- the relevant master 
integrals~\cite{xtri1,xtri2,AGO1,AGO2,AGO3,planarA,nonplanarA,planarB,nonplanarB,planarIR}
have been calculated in the past few years using a wide variety of
methods.

\subsection{Tree level results}
\label{subsec:treelevel}

At tree level, only $A_1^{(0)}$ is non-zero,
\begin{eqnarray}
A_{1}^{(0),[1]}(\sab,\sbc) &=& \frac{1}{2\sab},\nonumber \\
A_{1}^{(0),[2]}(\sab,\sbc) &=& -\frac{1}{N} ~A_{1}^{(0),[1]}(\sab,\sbc).
\end{eqnarray}
Using the $\dd$-dimensional form of Eq.~(\ref{eq:square}) together with the colour
sum of Eq.~(\ref{eq:CC}) we find that the self-interference of the tree amplitude 
is given by the usual result
\begin{equation}
\label{eq:squaretree}
\sum_{\rm spins} \langle {\cal M}^{(0)}|  {\cal M}^{(0)}\rangle  =
2 (N^2-1) ~\left(\frac{\sac^2+\sbc^2-\epsilon \sab^2}{\sab^2}\right). 
\end{equation}
Similarly the interference of the identical quark amplitudes is given by
\begin{equation}
\label{eq:squaretreeid}
\sum_{\rm spins} -\langle \overline{\cal M}^{(0)}|  {\cal M}^{(0)}\rangle -\langle {\cal M}^{(0)}|  \overline{\cal M}^{(0)}\rangle  =
-4 \left(\frac{N^2-1}{N}\right) (1-\ep)~\left(\frac{\sac^2}{\sab\sbc}+\ep\right). 
\end{equation}

\subsection{One-loop results}
\label{subsec:oneloop}

At one-loop, the first four Dirac structures
 ${\cal D}_1\ldots{\cal D}_4$ contribute and 
the unrenormalised coefficients can be expressed  
in terms of the one-loop
box integral in $\dd=6-2\epsilon$ dimensions, Box$^6(s_{ij},s_{ik})$, and the
one-loop bubble, Bub$(s_{ij})$.   
To all orders in $\epsilon = (4-\dd)/2$, we find that,
\begin{eqnarray}
A_1^{(1),[1]}(\sab,\sbc)=
&+&\frac{(-2 \sbc+\sab) \xac}{2 \sbc \sab N} +\frac{\xac}{N \sab \e} \\\nonumber
&+&\frac{(N^2-2) (-2 \sbc+\sab) \xbc}{4 \sbc \sab N} \\\nonumber
&+&\frac{(N^2-2) \xbc}{2 N \sab \e} \\\nonumber
&-&\frac{(-18 \e N^2+9 \e^2 N^2+6 N^2+7 \e-8 \e^2+4 \e^3-6) \Bab}
{4 N \sab (-3+2 \e) \e} \\\nonumber
&+&\frac{(\e-1) \NF \Bab}{2 \sab (-3+2 \e)} \\\nonumber
&-&\frac{(N^2-2) (5 \sab \sbc+2 \sbc^2+\sab^2) \e \Boxabbc}{4 \sbc \sab N} -\frac{(-2 \sbc^2+\sab^2) \e \Boxabac}{2 \sbc \sab N} \\\nonumber
&-&\frac{\sac (-2 \sbc+\sab) \Boxabac}{2 \sbc \sab N} +\frac{(N^2-2) (2 \sab+\sbc) \Boxabbc}{2 \sab N} \\\nonumber
A_2^{(1),[1]}(\sab,\sbc)=
&+&\frac{2 \xac}{\sbc \sab N} 
-\frac{(N^2-2) (2 \sbc+\sab) \xbc}{2 N \sab \sac \sbc} \\\nonumber
&-&\frac{(N^2-2) \e \xbc}{2 N \sac \sbc} +\frac{\sab (N^2-2) \e^2 \Boxabbc}{2 N \sac \sbc} \\\nonumber
&+&\frac{(N^2-2) (2 \sbc+\sab) (\sab-\sbc) \e \Boxabbc}{2 N \sab \sac \sbc} 
+\frac{2 \sac \e \Boxabac}{\sbc \sab N} \\\nonumber
&+&\frac{2 \Boxabac}{N \sbc} -\frac{(N^2-2) \Boxabbc}{2 N \sac} \\\nonumber
A_3^{(1),[1]}(\sab,\sbc)=
&+&\frac{\Boxabac}{8 N} -\frac{(N^2-2) \Boxabbc}{16 N} \\\nonumber
A_4^{(1),[1]}(\sab,\sbc)=
&-&\frac{\xac}{4 N \sac \sbc} -\frac{(N^2-2) \xbc}{8 N \sac \sbc} \\\nonumber
&+&\frac{\Boxabac}{4 N \sbc} +\frac{(N^2-2) \Boxabbc}{8 N \sac} \\\nonumber
&+&\frac{\sab (N^2-2) \e \Boxabbc}{8 N \sac \sbc} +\frac{\sab \e \Boxabac}{4 N \sac \sbc} \\\nonumber
A_1^{(1),[2]}(\sab,\sbc)=
&-&\frac{(N^2+1) (-2 \sbc+\sab) \xac}{4 N^2 \sab \sbc} \\\nonumber
&-&\frac{(N^2+1) \xac}{2 N^2 \e \sab} \\\nonumber
&+&\frac{(-2 \sbc+\sab) \xbc}{4 N^2 \sab \sbc} \\\nonumber
&+&\frac{\xbc}{2 N^2 \e \sab} \\\nonumber
&+&\frac{(-18 \e N^2+9 \e^2 N^2+6 N^2+7 \e-8 \e^2+4 \e^3-6) \Bab}{4 \sab (-3+2 \e) N^2 \e} \\\nonumber
&-&\frac{(\e-1) \NF \Bab}{2 N \sab (-3+2 \e)} \\\nonumber
&-&\frac{(5 \sab \sbc+2 \sbc^2+\sab^2) \e \Boxabbc}{4 N^2 \sab \sbc} +\frac{(N^2+1) (-2 \sbc^2+\sab^2) \e \Boxabac}{4 N^2 \sab \sbc} \\\nonumber
&+&\frac{\sac (N^2+1) (-2 \sbc+\sab) \Boxabac}{4 N^2 \sab \sbc} +\frac{(2 \sab+\sbc) \Boxabbc}{2 N^2 \sab} \\\nonumber
A_2^{(1),[2]}(\sab,\sbc)=
&-&\frac{(N^2+1) \xac}{N^2 \sab \sbc} -\frac{(2 \sbc+\sab) \xbc}{2 N^2 \sab \sac \sbc} \\\nonumber
&-&\frac{\e \xbc}{2 N^2 \sac \sbc} +\frac{\sab \e^2 \Boxabbc}{2 N^2 \sac \sbc} \\\nonumber
&-&\frac{(N^2+1) \Boxabac}{N^2 \sbc} 
-\frac{\Boxabbc}{2 N^2 \sac} \\\nonumber
&+&\frac{(2 \sbc+\sab) (\sab-\sbc) \e \Boxabbc}{2 N^2 \sab \sac \sbc} 
-\frac{(N^2+1) \sac \e \Boxabac}{N^2 \sab \sbc} \\\nonumber
A_3^{(1),[2]}(\sab,\sbc)=
&-&\frac{(N^2+1) \Boxabac}{16 N^2} -\frac{\Boxabbc}{16 N^2} \\\nonumber
A_4^{(1),[2]}(\sab,\sbc)=
&+&\frac{(N^2+1) \xac}{8 N^2 \sac \sbc} -\frac{\xbc}{8 N^2 \sac \sbc} \\\nonumber
&-&\frac{(N^2+1) \Boxabac}{8 N^2 \sbc} +\frac{\Boxabbc}{8 N^2 \sac} \\\nonumber
&+&\frac{\sab \e \Boxabbc}{8 N^2 \sac \sbc} -\frac{\sab (N^2+1) \e \Boxabac}{8 N^2 \sac \sbc}. \\\nonumber
\end{eqnarray}
Explicit expansions in $\epsilon$ for the $d=4-2\epsilon$ one-loop bubble and 
$d=6-2\epsilon$ one-loop box
graphs through to ${\cal O}(\ep^2)$ can be found, for example, in the Appendix of Ref.~\cite{qqgg}.

\subsection{Ultraviolet renormalisation}
\label{subsec:renorm}

The renormalisation of the matrix element in the \MSbar\ scheme is carried out by replacing 
the bare coupling $\alpha_0$ with the renormalised coupling 
$\alpha_s\equiv \alpha_s(\mu^2)$,
evaluated at the renormalisation scale $\mu^2$
according to Eq.~(\ref{eq:alpha}).

We denote the $i$-loop contribution to the unrenormalised tensor
coefficients  by 
$A_{I}^{(i),{\rm un}}$, using the same normalization as 
for the decomposition of the tensor coefficients in Eq.~(\ref{eq:perturb}).
The renormalised coefficients are then obtained as
\begin{eqnarray}
\label{eq:Aren}
A_{I}^{(0)}(\sab,\sbc)  &=& A_{I}^{(0),{\rm un}}(\sab,\sbc) ,
 \nonumber \\
A_{I}^{(1)}(\sab,\sbc)  &=& 
S_\e^{-1} A_{I}^{(1),{\rm un}}(\sab,\sbc) 
-\frac{\beta_0}{\e} A_{I}^{(0),{\rm un}}(\sab,\sbc)  ,  \nonumber \\
A_{I}^{(2)}(\sab,\sbc) &=& 
S_\e^{-2} A_{I}^{(2),{\rm un}} (\sab,\sbc) 
-\frac{2\beta_0}{\e} S_\e^{-1}
A_{I}^{(1),{\rm un}}(\sab,\sbc)  \nonumber \\
&&
-\frac{1}{2}\left(\frac{\beta_1}{\e}-\frac{2\beta_0^2}{\e^2}\right)
A_{I}^{(0),{\rm un}}(\sab,\sbc).
\end{eqnarray}

\subsection{Infrared behaviour of the tensor coefficients}
\label{subsec:infrared}

After performing ultraviolet renormalisation,
the amplitudes still
contain singularities which are of infrared origin and will be  analytically
cancelled by those occurring in radiative processes of the
same order.
It has become common practice to organize the 
infrared pole structure of the one- and two-loop contributions renormalised in the 
\MSbar\ scheme in terms of the tree and renormalised one-loop amplitudes multiplied by
singular operators according to Catani~\cite{catani}. 
Precisely the same procedure applies to the tensor coefficients.

In particular, the infrared behaviour of the one-loop coefficients is given by
\begin{eqnarray}
\label{eq:IR1L}
A_{I}^{(1)}(\sab,\sbc) &=& {\bom I}^{(1)}(\epsilon,\sab,\sbc) A_{I}^{(0)}(\sab,\sbc) + A_{I}^{(1),{\rm finite}}(\sab,\sbc),
\end{eqnarray}
while the two-loop singularity structure is
\begin{eqnarray}
A_{I}^{(2)}(\sab,\sbc) &=& {\bom I}^{(2)}(\epsilon,\sab,\sbc)  A_{I}^{(0)}(\sab,\sbc)
+ {\bom I}^{(1)}(\epsilon,\sab,\sbc) A_{I}^{(1)}(\sab,\sbc)
\nonumber \\
&&+ A_{I}^{(2),{\rm finite}}(\sab,\sbc),
\end{eqnarray}
where ${\bom I}^{(2)}(\epsilon,\sab,\sbc)$ is given by,
\begin{eqnarray}
\label{eq:IR2L}
{\bom I}^{(2)}(\epsilon,\sab,\sbc)&=&
-\frac{1}{2}  {\bom I}^{(1)}(\epsilon,\sab,\sbc) {\bom I}^{(1)}(\epsilon,\sab,\sbc)
-\frac{\beta_0}{\epsilon} {\bom I}^{(1)}(\epsilon,\sab,\sbc) \nonumber \\
&&
+e^{-\epsilon \gamma } \frac{ \Gamma(1-2\epsilon)}{\Gamma(1-\epsilon)} 
\left(\frac{\beta_0}{\epsilon} + K\right)
{\bom I}^{(1)}(2\epsilon,\sab,\sbc)  + {\bom H}^{(2)}(\epsilon,\sab,\sbc) 
\end{eqnarray}
and the constant $K$ is
\begin{equation}
K = \left( \frac{67}{18} - \frac{\pi^2}{6} \right) \CA - 
\frac{10}{9} T_R \NF.
\end{equation}
The implicit dependence of ${\bom I^{(1)}}$, ${\bom I^{(2)}}$ and 
${\bom H^{(2)}}$ on $\sac=-\sab-\sbc$ has been suppressed.

In QCD the singular ${\bom I}^{(1)}(\epsilon)$ operator is a matrix in colour space and 
is given by~\cite{catani}
\begin{eqnarray}
{\bom I^{(1)}}(\ep) &=& \frac{1}{2}\frac{e^{\ep \gamma }}{\Gamma(1-\ep)}
     \sum_i \nu^{sing}_i(\ep) \sum_{j \neq i} {\bf T}_i
   \cdot {\bf T}_j \left(-\frac{\mu^2}{2p_i \cdot p_j}
\right)^{\epsilon}
\end{eqnarray}
where the sum runs over the pairs of external coloured particles involved in the process and
the singular function is
\begin{equation}
\nu^{sing}_i(\ep) = \frac{1}{\ep^2} + \gamma_i \frac{1}{\ep},\qquad
\gamma_{q,\bar q}=\frac{3}{2}, \qquad \gamma_g=\frac{\beta_0}{\CA}
\end{equation}
As usual, the colour charge algebra is defined as
$$
{\bf T}_i \cdot {\bf T}_j = \left\{ \begin{array}{ll}
                                   {\bf T}_j \cdot {\bf T}_i & \mbox{if $i
                                     \neq j$,} \\
                                   {\bf T}_i^2 = C_i & \mbox{otherwise,}
                                   \end{array} \right.
$$
with the Casimir $C_i =C_F$ ($C_i =C_A$) if parton $i$ is a quark (gluon) and where
the colour charge ${\bf T}_i$ is  $t^a_{cb}$ ($-t^a_{cb}$) for a quark (anti-quark) and $if_{cab}$
for a gluon.

For the colour basis for the four-quark process presented in Eq.~(\ref{eq:colour}), we have the 
following operator 
\begin{eqnarray}
\label{eq:opIgg}
\bom{I}^{(1)}(\ep,\sab,\sbc) = -\frac{e^{\ep\gamma}}{\Gamma(1-\ep)}\Biggl({1\over \ep^2}+{3\over 2  \ep}\Biggr)
\left(
\begin{array}{cc}
{\bom A}(\ep,\sab,\sbc)  & ~{\bom B}(\ep,\sab,\sbc) \\
{\bom B}(\ep,\sbc,\sab) & ~{\bom A}(\ep,\sbc,\sab) \\
\end{array}
\right) \nonumber \\
\end{eqnarray}
where
\begin{eqnarray}
\label{eq:Adefqg}
{\bom A}(\ep,\sab,\sbc) &=&
\frac{(N^2-1)}{N}{\ft}+\frac{1}{N}\left[{\fu}-{\fs}\right],\\
\label{eq:Bdefqg}
{\bom B}(\ep,\sab,\sbc) &=&
\left[{\ft}-{\fu}\right].
\end{eqnarray}

In Eq.~(\ref{eq:IR2L}) the function ${\bom H}^{(2)}$ contains poles of
$\O{1/\ep}$ and is process and renormalisation scheme dependent. The specific
structure of this term  was not given in Ref.~\cite{catani} for the general
case, but in the case where a single $q \bar q$ pair is involved,  ${\bom
H}^{(2)}$ is related to the quark electromagnetic form factor. Subsequent work
by Sterman and Tejeda-Yeomans~\cite{stermanTY} has shown that the 
form factor is
in fact the backbone of the singularity  structure of multi-loop QCD
amplitudes. 

The colour uncorrelated part of the ${\bom H}^{(2)}$ function has now been
established for all  $2 \to 2$ partonic processes by 
direct Feynman diagram evaluation of two-loop matrix
elements~\cite{BDGbha,qqunlike,qqlike,qqgg,gggg,qqpp}.  More precisely, it was
found that each external coloured leg in the  partonic process contributes
independently to the matrix element given by
\begin{eqnarray}
\bra{\cm_0} {\bom H}^{(2)} \ket{\cm_0} = \frac{e^{\ep\gamma}}{4 \ep \Gamma(1-\ep)}
H^{(2)}\braket{\cm_0}{\cm_0}.
\end{eqnarray}
For the processes we are interested in here, 
\begin{eqnarray}
\label{eq:nqng}
H^{(2)} = 2 H_{q}^{(2)} + 2 H_{\bar q}^{(2)}
\end{eqnarray}
with
\begin{eqnarray}
H_{q}^{(2)} =H_{\bar q}^{(2)} &=&\left(\frac{\pi^2}{2}-6 ~\zeta_3 
-\frac{3}{8}\right) \CF^2
+\left(\frac{13}{2}\zeta_3 +\frac{245}{216}-\frac{23}{48} \pi^2 \right) \CA \CF
\nonumber \\
&& + \left(-\frac{25}{54}+\frac{\pi^2}{12} \right) \TR \NF \CF.
\end{eqnarray}

However, ${\bom H}^{(2)}$  also contains additional non-trivial colour
correlations~\cite{catani} which can be explored  by calculating the hard scattering
amplitude, rather than the interference of two-loop and tree amplitudes where
the sum over colours causes the colour correlations to vanish.  Such colour
correlations are independent of the helicity configuration and have been
identified in the two-loop helicity amplitudes for the four gluon and the
two-quark two-gluon interactions which both involve four coloured
particles~\cite{BFDgggg,BFDqqgg,qqgghel}. According to the analysis of
Ref.~\cite{stermanTY} the colour-correlated contributions to  ${\bf \hat
H}^{(2)}$ are due to soft gluons\footnote{Note that these type of  colour
correlations between three different partons also appeared in the analysis of
higher-order  contributions to the soft-gluon current in Ref.~\cite{sone3}.}
and are directly related to soft anomalous  dimension matrices obtained by
studying the evolution of colour exchange in QCD hard  scattering (see
~\cite{KOS} and references therein). They are therefore also independent of
the type of (coloured) external particle.

For the four-quark processes studied here, we find that the colour correlations are 
indeed helicity independent
and are given by
\begin{eqnarray}\label{eq:h2coloura}
{\bom H}^{(2)}(\ep,\sab,\sbc) = \frac{e^{\ep\gamma}}{4 \ep
\Gamma(1-\ep)}\left[{\fsd}+\ftd-\fud\right]
\Biggl( H^{(2)}~{\bom  1} + {\bf \widehat H}^{(2)}(\sab,\sbc) \Biggr)\nonumber \\
\end{eqnarray}
with $H^{(2)}$ as given in Eq.~(\ref{eq:nqng}) and
\begin{eqnarray}\label{eq:h2colourb}
{\bf \widehat H}^{(2)}(\sab,\sbc) &=&
-4~i  ~f_{abc} ~{\bom T}^{a[1]} {\bom T}^{b[2]} {\bom T}^{c[3]}
~{\rm ln}\left(\frac{-\sab}{-\sbc} \right) 
{\rm ln}\left(\frac{-\sbc}{-\sac} \right) 
{\rm ln}\left(\frac{-\sac}{-\sab} \right).
\end{eqnarray}
The structure of ${\bf \widehat H}^{(2)}(\sab,\sbc)$  is in complete agreement
with the explicit results found in Ref.~\cite{BFDgggg,BFDqqgg,qqgghel}  and
confirms that the colour correlations are independent of the external
particles.    Note that the square bracket in Eq.~(\ref{eq:h2coloura}) is, to
some extent, arbitrary and other choices may be equally valid.   We have made
this particular choice partly to match the choice made in
Refs.~\cite{qqunlike,qqlike}, but also to ensure that no additional 
finite contribution from the colour correlation term is 
generated when the "squared" matrix elements for identical quark scattering are
reconstructed.\footnote{Colour automatically kills the contribution from 
${\bf \widehat H}^{(2)}$ when the quarks are distinct.}  

Note that the matrix $f_{abc} ~{\bom T}^{a[1]} {\bom T}^{b[2]} {\bom T}^{c[3]}$ collects all 
colour weights that result from the analysis of triple gluon vertices connecting any three 
different external partons, projected into the colour basis given in Eq.~(\ref{eq:colour}), 
so that for the four-quark process,
\begin{eqnarray}
i~f_{abc} ~{\bom T}^{a[1]} {\bom T}^{b[2]} {\bom T}^{c[3]} = 
-\frac{1}{4}
\left(
\begin{array}{cc}
-1 & -N  \\
N & 1
\end{array}
\right).
\end{eqnarray}

\section{Helicity amplitudes}
\label{sec:helamp}

The helicity amplitudes $\M_{\lambda_1\lambda_2\lambda_3\lambda_4}$ 
can be obtained from the general $d$-dimensional tensor of
Eq.~(\ref{eq:gentensor}) by setting the dimensionality of the Lorentz matrices
to be four and 
using standard four-dimensional helicity techniques.  This  corresponds to working in
the 't Hooft-Veltman scheme.  We use the standard convention of denoting
the two helicity states of a
four-dimensional light-like spinor $\psi(p)$  by,
\begin{equation}
\psi_\pm(p)= \frac{1}{2}(1\pm \gamma_5)\psi(p),
\end{equation}
with the further notation,
\begin{equation}
|p\pm\rangle = \psi_\pm(p), \qquad\qquad
\langle p\pm | = \overline{\psi_\pm(p)}.
\end{equation}
Particles may thus be crossed to the initial state by reversing the sign of the
helicity.
The basic quantity is the spinor product,
\begin{equation}
\langle pq\rangle = \langle p- | q+\rangle, \qquad\qquad
[ pq]= [ p+ | q-],
\end{equation}
such that 
\begin{equation}
\langle pq\rangle  [qp] = 2 p.q.
\end{equation}

Fixing the helicity of the (final-state) quark $q$ to be positive, $\lambda_1 = +$ (and therefore
the helicity of the antiquark $\bar q$ is negative, $\lambda_2 = -$) 
 we find that, 
 up to a helicity dependent phase, the two independent helicity amplitudes are given by
\begin{eqnarray}
\label{eq:helamp}
\M_{+-+-} &\propto& \sac \left( B_1(\sab,\sbc)+\delta_{qQ} B_1(\sbc,\sab)\right),\nonumber \\
%\biggl(2A_1(\sab,\sbc)-\sbc A_2(\sab,\sbc)+32 A_3(\sab,\sbc)+4\sac A_4(\sab,\sbc)\nonumber \\
%&& \qquad\qquad
%+512 A_5(\sab,\sbc) + 64\sac A_6(\sab,\sbc)\biggr),\nonumber \\
\M_{+--+} &\propto& 
\sbc B_2(\sab,\sbc),
%\biggl( 2A_1(\sab,\sbc)+\sac A_2(\sab,\sbc)+8 A_3(\sab,\sbc) +32 A_5(\sab,\sbc) \nonumber \\
%&& \qquad\qquad
%+ 16\sac A_6(\sab,\sbc)\biggr).
\end{eqnarray}
where the combinations of tensor coefficients $B_1$ and $B_2$ are defined in Eq.~(\ref{eq:b1andb2}).
In addition, when the quarks are identical, there is the additional helicity amplitude,
\begin{equation}
\M_{++--} \propto 
\delta_{qQ}~\sab B_2(\sbc,\sab).
\end{equation}
Helicity amplitudes where the quark helicity is negative are obtained through parity,
\begin{equation}
\M_{--\lambda_2-\lambda_3-\lambda_4} 
= 
\left(\M_{+\lambda_2\lambda_3\lambda_4}\right)^*.
\end{equation}
The helicity amplitudes are functions of the external scales, however the 
dependence on $\sab$ and $\sbc$ (together with $\sac=-\sab-\sbc$) has been suppressed.

By inspection, we see that ``squaring'' and summing over the helicities generates the
interference of amplitudes given in Eqs.~(\ref{eq:squaretree}) and (\ref{eq:squaretreeid}). Thus, the helicity
amplitudes can be obtained directly from a Feynman diagram calculation by projecting
out the individual tensor coefficients and taking the appropriate linear combination.

Just as for the individual tensor coefficients, the helicity amplitudes 
can be perturbatively decomposed as,
\begin{eqnarray}
\lefteqn{|{\cal M}\rangle_{\lambda_1\lambda_2\lambda_3\lambda_4}}\nonumber \\
 &=&   4\pi \,\alpha_s \left[
|{\cal M}^{(0)}\rangle_{\lambda_1\lambda_2\lambda_3\lambda_4}  
+ \left(\frac{\alpha_s}{2\pi}\right) |{\cal M}^{(1)}\rangle_{\lambda_1\lambda_2\lambda_3\lambda_4} 
+ \left(\frac{\alpha_s}{2\pi}\right)^2 |{\cal M}^{(2)}\rangle_{\lambda_1\lambda_2\lambda_3\lambda_4}
+ {\cal O}(\alpha_s^3) \right].
\end{eqnarray}
As in Eq.~(\ref{eq:Aren})  
the renormalised helicity amplitudes are obtained from the unrenormalised
amplitudes by 
\begin{eqnarray}
\label{eq:helren}
|{\cal M}^{(0)}\rangle_{\lambda_1\lambda_2\lambda_3\lambda_4}
  &=& |{\cal M}^{(0),{\rm un}}\rangle_{\lambda_1\lambda_2\lambda_3\lambda_4},
 \nonumber \\
|{\cal M}^{(1)}\rangle_{\lambda_1\lambda_2\lambda_3\lambda_4} &=& 
S_\e^{-1} |{\cal M}^{(1),{\rm un}}\rangle_{\lambda_1\lambda_2\lambda_3\lambda_4} 
-\frac{\beta_0}{\e} |{\cal M}^{(0),{\rm un}}\rangle_{\lambda_1\lambda_2\lambda_3\lambda_4}  ,  \nonumber \\
|{\cal M}^{(2)}\rangle_{\lambda_1\lambda_2\lambda_3\lambda_4} &=& 
S_\e^{-2} |{\cal M}^{(2),{\rm un}}\rangle_{\lambda_1\lambda_2\lambda_3\lambda_4}  
-\frac{2\beta_0}{\e} S_\e^{-1}
|{\cal M}^{(1),{\rm un}}\rangle_{\lambda_1\lambda_2\lambda_3\lambda_4}  \nonumber \\
&&
-\frac{1}{2}\left(\frac{\beta_1}{\e}-\frac{2\beta_0^2}{\e^2}\right)
|{\cal M}^{(0),{\rm un}}\rangle_{\lambda_1\lambda_2\lambda_3\lambda_4},\nonumber \\
\end{eqnarray}
while the infrared singularity structure is given by,
\begin{eqnarray}
\label{eq:remainder1}
|{\cal M}^{(1)}\rangle_{\lambda_1\lambda_2\lambda_3\lambda_4} &=& 
{\bom I}^{(1)}(\ep,\sab,\sbc) |{\cal M}^{(0)}\rangle_{\lambda_1\lambda_2\lambda_3\lambda_4} 
+|{\cal M}^{(1),{\rm finite}}\rangle_{\lambda_1\lambda_2\lambda_3\lambda_4},\\
\label{eq:remainder2}
|{\cal M}^{(2)}\rangle_{\lambda_1\lambda_2\lambda_3\lambda_4} &=& 
{\bom I}^{(2)}(\ep,\sab,\sbc)  
|{\cal M}^{(0)}\rangle_{\lambda_1\lambda_2\lambda_3\lambda_4}
+ {\bom I}^{(1)}(\ep,\sab,\sbc) 
|{\cal M}^{(1)}\rangle_{\lambda_1\lambda_2\lambda_3\lambda_4}\nonumber \\
&&+ |{\cal M}^{(2),{\rm finite}}\rangle_{\lambda_1\lambda_2\lambda_3\lambda_4}, 
\end{eqnarray}
where the operators ${\bom I}^{(i)}$ are the process dependent matrices in 
colour space given in Section~\ref{subsec:infrared}.

Similarly, the $n$-loop helicity amplitudes are vectors in colour space and can be further
decomposed as in Eq.~(\ref{eq:colour}),
\begin{eqnarray}
\label{eq:helcolour}
|{\cal M}^{(n)}\rangle_{\lambda_1\lambda_2\lambda_3\lambda_4} &=&  
\sum_{i=1}^2 \;\C_i\;|{\cal M}^{(n),[i]}\rangle_{\lambda_1\lambda_2\lambda_3\lambda_4}.
\end{eqnarray}

\subsection{Helicity amplitudes for physical processes}
\label{sec:results}
The physically relevant scattering amplitudes 
we wish to describe are the unlike quark processes,
\begin{eqnarray}
\label{eq:schannel}
s: \qquad\qquad&&q(p_2,+) + \bar q(p_1,-)         \to  Q(p_3,\lambda_3)+\bar Q (p_4,\lambda_4),\\
\label{eq:tchannel1}
t: \qquad\qquad&&q(p_2,+) + \bar Q(p_1,\lambda_1) \to  q(p_3,+)        +\bar Q(p_4,\lambda_4),\\
\label{eq:tchannel2}
t^\prime: \qquad\qquad&&q(p_2,+) +      Q(p_1,\lambda_1) \to  q(p_3,+)        +     Q(p_4,\lambda_4),\\
\label{eq:uchannel}
u: \qquad\qquad&&q(p_2,+) +      Q(p_1,\lambda_1) \to  Q(p_3,\lambda_3)        +     q(p_4,+).
\end{eqnarray}
For convenience, we denote each of the processes of Eqs.~(\ref{eq:schannel})--(\ref{eq:uchannel})
as belonging to a particular channel (e.g. Eq.~(\ref{eq:schannel}) is the $s$-channel and so on)
where  $s$, $t$ and $u$ are the usual Mandelstam variables,
$s = s_{12} > 0$, 
$t = s_{23} < 0$, $u = s_{13} < 0$ and $s+t+u = 0$.
Processes where the initial state quark has negative helicity are obtained by a
parity transformation, while processes with $\bar q(p_2)$ in the initial state are
obtained by charge conjugation.
Note that the $t^\prime$-channel amplitudes can be obtained by exchanging $1
\leftrightarrow 4$ in the $t$-channel amplitudes or by exchanging $3 \leftrightarrow 4$
in the $u$-channel amplitudes.

To present the helicity amplitudes for the various processes, 
it is convenient to organise the amplitude where particles 1 and 2 are in the initial state
in terms of a spinor factor ${\cal S}$ and the colour factor $\C$  so that,
\begin{equation}
\label{eq:Mdecomp}
|{\cal M}_c^{(n)}\rangle_{\lambda_2\lambda_1\lambda_3\lambda_4} =
\sum_i ~{\cal S}^{[i]}_{c\lambda_2\lambda_1\lambda_3\lambda_4}  \times 
\C_{ci}\times
|\widehat {\cal M}_c^{(n),[i]}\rangle_{\lambda_2\lambda_1\lambda_3\lambda_4},
\end{equation}
where $c$ (= $s$, $t$, $u$) denotes the channel and 
the sum runs over the two colour structures.
Explicitly, we have 
\begin{equation}
\label{eq:scolour}
\C_{s1} = {\bom \delta_{i_3i_2}}{\bom \delta_{i_1i_4}}, \qquad
\C_{s2} = {\bom \delta_{i_3i_4}}{\bom \delta_{i_1i_2}}. \qquad
\end{equation}
The $u$- and $t$-channel colour factors are obtained by crossing symmetry,
\begin{equation}
\C_{ti} = \C_{si} ~~(1 \leftrightarrow 3),\qquad\qquad
\C_{ui} = \C_{si} ~~(1 \leftrightarrow 4),
\end{equation}
while the $t^\prime$-channel colour factor is given by,
\begin{equation}
\C_{t^\prime i} =\C_{ti}(1 \leftrightarrow 4) = \C_{ui}(3 \leftrightarrow 4).
\end{equation}
We note that $\C_{t1} =\C_{s2}$, $\C_{t2}=\C_{s1}$ and
$\C_{t^\prime 1} =\C_{u2}$, $\C_{t^\prime 2}=\C_{u1}$.

There is considerable freedom in the definition of the helicity dependent spinor factors.
For the $s$-channel, for process~(\ref{eq:schannel}) we choose,
\begin{eqnarray}
\label{eq:s+-+-}
{\cal S}_{s+-+-}^{[i]} &=& -i\frac{[13]\langle 42\rangle}
{s_{12}},\nonumber \\
{\cal S}_{s+--+}^{[i]} &=& -i\frac{[14]\langle 32\rangle}
{s_{12}}.
\end{eqnarray}
The spinor prefactors for the other channels are also obtained
by crossing symmetry,
\begin{equation}
{\cal S}_{t\lambda_2\lambda_1\lambda3\lambda_4}^{[i]} = 
{\cal S}_{s\lambda_2-\lambda_3-\lambda_1\lambda_4}^{[i]} ~~(p_1 \leftrightarrow p_3),\qquad\qquad
{\cal S}_{u\lambda_2\lambda_1\lambda_3\lambda_4}^{[i]} = 
{\cal S}_{s\lambda_2-\lambda_4\lambda_3-\lambda_1}^{[i]} ~~(p_1 \leftrightarrow p_4),
\end{equation}
and,
\begin{equation}
{\cal S}_{t^\prime\lambda_2\lambda_3\lambda_4\lambda_4}^{[i]} = 
{\cal S}_{u\lambda_2\lambda_1\lambda_4\lambda_3}^{[i]} ~~(p_3 \leftrightarrow p_4).
\end{equation}

The helicity amplitudes for the like-quark scattering processes,
\begin{eqnarray}
\label{eq:stchannel}
st: \qquad\qquad&&q(p_2,+) + \bar q(p_1,-)         \to  q(p_3,\lambda_3)+\bar q (p_4,\lambda_4),\\
\label{eq:utchannel}
ut^\prime: \qquad\qquad&&q(p_2,+) +      q(p_1,\lambda_1) \to  q(p_3,+)        +     q(p_4,\lambda_4),
\end{eqnarray}
are derived directly from the unlike quark amplitudes.
Here we have labelled the two processes according to the two channels that contribute.
Explicitly, we have
\begin{eqnarray}
|{\cal M}_{st}^{(n)}\rangle_{+-+-} &=& \phantom{-}|{\cal M}_s^{(n)}\rangle_{+-+-} -|{\cal M}_t^{(n)}\rangle_{+-+-},\\
|{\cal M}_{st}^{(n)}\rangle_{+--+} &=& \phantom{-}|{\cal M}_s^{(n)}\rangle_{+-+-} ,\\
|{\cal M}_{st}^{(n)}\rangle_{++++} &=&  -|{\cal M}_t^{(n)}\rangle_{++++},\\
|{\cal M}_{ut^\prime}^{(n)}\rangle_{++++} &=& \phantom{-}|{\cal M}_u^{(n)}\rangle_{++++} -|{\cal M}_{t^\prime}^{(n)}\rangle_{++++},\\
|{\cal M}_{ut^\prime}^{(n)}\rangle_{+--+} &=& \phantom{-}|{\cal M}_u^{(n)}\rangle_{+--+},\\
|{\cal M}_{ut^\prime}^{(n)}\rangle_{+-+-} &=&  -|{\cal M}_{t^\prime}^{(n)}\rangle_{+-+-},
\end{eqnarray}
where the minus sign is due to the exchange of identical particles.

\subsection{Tree level helicity amplitudes}

At tree-level, inserting the results for the 
tensor coefficients given in Section~\ref{subsec:treelevel},
we find the only non-vanishing helicity amplitudes in the $s$-channel are,
\begin{eqnarray}
 |\widehat {\cal M}_s^{(0),[1]}\rangle_{+-+-} = |\widehat {\cal M}_s^{(0),[1]}\rangle_{+--+} &=& 1,\nonumber\\
 |\widehat {\cal M}_s^{(0),[2]}\rangle_{+-+-} = |\widehat {\cal M}_s^{(0),[2]}\rangle_{+--+}&=& -\frac{1}{N}.
\end{eqnarray}
The appropriate spinor factors are obtained from Eq.~(\ref{eq:s+-+-}) 
while the colour factors are given by 
Eq.~(\ref{eq:scolour}).  Tree helicity amplitudes for the other channels are
obtained by the appropriate crossing symmetry.

\subsection{One-loop helicity amplitudes}
\label{subsec:oneloopamps}

The one-loop helicity amplitudes are straightforwardly obtained by inserting the all-orders results for 
the tensor coefficients given in Section~\ref{subsec:oneloop} into Eq.~(\ref{eq:helamp}),
expanding the one-loop bubble and box
integrals around $\epsilon = 0$, and renormalising according to Eq.~(\ref{eq:helren}).
Explicit expansions in $\epsilon$ for the $d=4-2\epsilon$ one-loop bubble and $d=6-2\epsilon$ one-loop box
graphs can be found, for example, in the Appendix of Ref.~\cite{qqgg}.
   
The finite remainders of the one-loop amplitudes defined through Eq.~(\ref{eq:remainder1})
for the distinct quark scattering process can be 
decomposed according to the number of colours and massless quark flavours,
\begin{eqnarray}
|\widehat {\cal M}_c^{(1),[1],finite}\rangle
&=& N A_c^{(1),[1]} + \frac{1}{N} B_c^{(1),[1]} + \NF
C_c^{(1),[1]}-\beta_0\Ls\,|\widehat {\cal M}_c^{(0),[1]}\rangle,\nonumber \\
|\widehat {\cal M}_c^{(1),[2],finite}\rangle
&=& A_c^{(1),[2]} + \frac{1}{N^2} B_c^{(1),[2]} + \frac{\NF}{N}
C_c^{(1),[2]}-\beta_0\Ls\,|\widehat {\cal M}_c^{(0),[2]}\rangle,
\end{eqnarray}
where $\beta_0$ is defined in Eq.~(\ref{betas}).
For clarity, the dependence on the helicities has been suppressed.  
Explicit expressions for $A,\ldots,C$
in the physical region, $s = s_{12} > 0$, 
$t = s_{23} < 0$ and $u = s_{13} < 0$, are given in Appendix~\ref{app:oneloop}.
We have checked that the finite one-loop helicity amplitudes presented here are in 
agreement with those given in Ref.~\cite{KST}.

\subsection{Two-loop helicity amplitudes}
\label{subsec:twoloopamps}

The main results of this paper are the two-loop amplitudes for processes
(\ref{eq:schannel})--(\ref{eq:uchannel}). As in the tree and one-loop cases,
the helicity amplitudes can be directly extracted from the appropriate linear
combination, Eq.~(\ref{eq:helamp}), of unrenormalised two-loop tensor
coefficients obtained by direct evaluation of the projectors given in
Section~\ref{subsec:projectors} acting on two-loop Feynman diagrams. 
Renormalisation is
achieved via  Eq.~(\ref{eq:helren}) and the two-loop master integrals expanded
around $\epsilon=0$. The finite remainder of the two-loop amplitudes are
defined through  Eq.~(\ref{eq:remainder2}) and, for each process can be 
decomposed according to the number of colours and massless quark flavours,
\begin{eqnarray}
|\widehat {\cal M}_c^{(2),[1],finite}\rangle
&=& N^2 A_c^{(2),[1]} +  B_c^{(2),[1]} + 
\frac{1}{N^2} C_c^{(2),[1]}
\nonumber \\
&& 
+N\NF D_c^{(2),[1]}
+  \frac{\NF}{N} E_c^{(2),[1]}
+ \NF^2 F_c^{(2),[1]} 
\nonumber\\
&&-2\beta_0\Ls\,|\widehat {\cal M}_c^{(1),[1],finite}\rangle
-(\beta_1 \Ls + \beta_0^2\Ls^2)\,|\widehat {\cal M}_c^{(0),[1]}\rangle,\\
|\widehat {\cal M}_c^{(2),[2],finite}\rangle
&=& N A_c^{(2),[2]} +  \frac{1}{N}B_c^{(2),[2]} + 
\frac{1}{N^3} C_c^{(2),[2]}
\nonumber \\
&& 
+\NF D_c^{(2),[2]}
+  \frac{\NF}{N^2} E_c^{(2),[2]}
+ \frac{\NF^2}{N} F_c^{(2),[2]} 
\nonumber\\
&&-2\beta_0\Ls\,|\widehat {\cal M}_c^{(1),[2],finite}\rangle
-(\beta_1 \Ls + \beta_0^2\Ls^2)\,|\widehat {\cal M}_c^{(0),[2]}\rangle.
\end{eqnarray}
As in the previous section, the dependence on the helicities has been suppressed.  
Explicit expressions for $A,\ldots,E$
for the various  processes in the physical region, $s = s_{12} > 0$, 
$t = s_{23} < 0$ and $u = s_{13} < 0$, are given in Appendix~\ref{app:twoloop}.

\section{Comparison with previous results}
\label{sec:matelem}

In recovering the square, or interference, of amplitudes in a particular channel $c$,
the contraction of the $n$-loop colour vector $|{\cal M}^{(n)}\rangle$ 
with a conjugate $m$-loop
colour vector $\langle {\cal M}^{(m)}| $ obeys the rule
\begin{eqnarray}
\label{eq:makesquare}
\langle {\cal M}^{(m)} | {\cal M}^{(n)} \rangle  
&=&
\sum_{\rm helicities} ~\sum_{\rm colours}
 |{\cal M}_c^{(n)}\rangle_{\lambda_1\lambda_2\lambda_3\lambda_4}^*
  \, |{\cal M}_c^{(n)}\rangle_{\lambda_1\lambda_2\lambda_3\lambda_4} 
  \,
\end{eqnarray}
where   $|{\cal M}_c^{(n)}\rangle_{\lambda_1\lambda_2\lambda_3\lambda_4} $ contains all of the spinor
and colour information for a particular helicity.
Expanding the colour and spinor factors $\C$ and ${\cal S}$, and 
dropping the explicit dependence on the helicities, we see that,
\begin{eqnarray}
\label{eq:makesquare2}
\langle {\cal M}^{(m)} | {\cal M}^{(n)} \rangle  
&=&
\sum_{\rm helicities} ~\sum_{\rm colours}
 ~\sum_{i,j} \C_{ci}^{*}\, \C_{cj} \; {\cal S}_c^{[i] *} \;{\cal S}_c^{[j]} \;
     |\widehat {\cal M}_c^{(n),[i]}\rangle^*
  \, |\widehat {\cal M}_c^{(n),[j]}\rangle 
  \, \nonumber \\
&=&  
\sum_{\rm helicities}  
 ~\sum_{i,j}   \, {\cal C\!C}_{ij} \; {\cal S\!S}_{ij} \; |\widehat {\cal M}_c^{(n),[i]}\rangle^*
  \, |\widehat {\cal M}_c^{(n),[j]}\rangle 
,
\end{eqnarray}
where the colour sum matrix ${\cal C\!C}$ is defined in Eq.~(\ref{eq:CC}).

In the $s$-channel, the helicity dependent spinor matrices 
${\cal S\!S}_c = {\cal S}_c^{[i] *} \;{\cal S}_c^{[j]}$ are
given by,
\begin{equation}
{\cal S\!S}_{s+-+-} = \frac{u^2}{s^2}\left(
\begin{array}{ccc}
1 & 0   \\
0 & 1    
\end{array}
\),
\qquad 
{\cal S\!S}_{s+--+} = \frac{t^2}{s^2}\left(
\begin{array}{ccc}
1 & 0   \\
0 & 1    
\end{array}
\).
\end{equation}
Similar matrices for the other helicities are obtained by a parity transformation,
while corresponding matrices for the $u$ and $t$-channels are obtained by crossing symmetry,
\begin{equation}
{\cal S\!S}_u = {\cal S\!S}_s (s \leftrightarrow t),
\qquad
{\cal S\!S}_t = {\cal S\!S}_s (s \leftrightarrow u),
\end{equation}
and
\begin{equation}
{\cal S\!S}_{t^\prime} = {\cal S\!S}_u (u \leftrightarrow t).
\end{equation}
 
Eqs.~(\ref{eq:makesquare}) and (\ref{eq:makesquare2}) can be used to recreate the full interference of
tree and two-loop amplitudes. However, we notice that we can immediately separate the singularities from
the finite parts using Eq.~(\ref{eq:remainder2}). The finite remainder, denoted by ${\cal
F}_{inite}^{2\times 0}(s,t,u)$
in Refs.~\cite{qqunlike,qqlike} is thus given by,
\begin{equation}
2 \Re \sum_{\rm helicities}  
 ~\sum_{i,j}   \, {\cal C\!C}_{ij} \; {\cal S\!S}_{ij} \; 
 |\widehat {\cal M}_c^{(0),[i]}\rangle^*
  \, |\widehat {\cal M}_c^{(2),[j],finite}\rangle .
\end{equation}
Note that by using helicities, 
we are implicitly treating the external particle states in
4-dimensions - the HV scheme.    However, the singular contributions to the
renormalised amplitudes defined in Eqs.~(\ref{eq:remainder1}), (\ref{eq:remainder2}) are given
relative to the tree amplitude and therefore any additional terms that would be produced in
CDR are automatically removed.  Treating the external
states differently for both the $n$-loop and tree amplitudes does not alter the finite
contribution.  Direct application of Eq.~(\ref{eq:makesquare}) for the real part of the tree and
two-loop interference ($n=0$ and $m=2$) in the HV scheme therefore generates the same finite parts given in
section 4.2 of Ref.~\cite{qqunlike} for  unlike quark-quark scattering 
and in section 3.2 of Ref.~\cite{qqlike}  for like quark-quark scattering obtained using CDR.

Similarly, the square of one-loop graphs is obtained using $n=1$ and $m=1$ in
Eq.~(\ref{eq:makesquare}). In principle, the one-loop amplitude should be expanded through to ${\cal
O}(\epsilon^2)$.   However, as noted in \cite{qqgg,BFDgggg,metythesis}, the finite self interference only contains
logarithms (up to the fourth power), but does not contain triple and quartic polylogs.  These terms
naturally arise when the ${\cal O}(\epsilon)$ and ${\cal O}(\epsilon^2)$ terms in the expansion of the
one-loop box are multiplied by singularities from the conjugate amplitude.  Such terms appear in the
Catani pole structure, $\langle {\cal M}^{(1)} | \bom I^{(1)} | {\cal M}^{(0)}\rangle$ where $| {\cal
M}^{(1)}\rangle$ does need to be expanded through to ${\cal O}(\epsilon^2)$. 
The finite parts of the renormalised one-loop self
interference remaining after Catani's prediction for the singularities has been removed should therefore
be
identical to those obtained by  summing the finite parts using Eq.~(\ref{eq:makesquare}).   
In this instance, the finite remainder denoted by ${\cal
F}_{inite}^{1\times 1}(s,t,u)$ in \cite{qqpp} is obtained from
\begin{equation}
\sum_{\rm helicities}  
 ~\sum_{i,j}   \, {\cal C\!C}_{ij} \; {\cal S\!S}_{ij} \; 
 |\widehat {\cal
 M}_c^{(1),[i],finite}\rangle^*
  \, |\widehat {\cal M}_c^{(1),[j],finite}\rangle .
\end{equation}
Unfortunately, the results of Ref.~\cite{qqqq1Lsq} are not presented in quite the same way and a direct
comparison is more difficult.  However, the coefficients of the
finite remainder for the one-loop squared amplitudes using the Catani pole structure
are presented in Ref.~\cite{metythesis} and
we have checked that one-loop square reconstructed from the helicity amplitudes (the HV scheme)
correctly reproduces the coefficients given in 
Appendix C of Ref.~\cite{metythesis}.

\section{Conclusions}
\label{sec:conclusions}

In this paper we have presented the analytic expressions for the one- and
two-loop helicity amplitudes for massless quark-quark scattering $q\bar q \to
Q\bar Q $ and those processes related by crossing symmetry. These amplitudes
were obtained by the construction of $d$-dimensional projection operators that
extract the coefficients of the most general tensor representation for the
amplitude, order by order in perturbation theory. Once the renormalised tensor
coefficients are known, the renormalised helicity amplitudes can be constructed
straightforwardly as a linear combination of the coefficients.    

We applied the projection
operators directly to the Feynman diagrams that contribute at each order in
$\as$. The projectors saturate the tensor structure of the Feynman diagram and
yield a set of scalar integrals that can be related to a basis set of master
integrals with the application of widely used reduction algorithms.  In fact,
we are  using exactly the same tools as we did in~\cite{qqunlike, qqlike},
except in that case, the projector was merely the conjugate tree amplitude.  
Because the projectors exist in $d$-dimensions, conventional dimensional
regularisation is preserved and there is no ambiguity in dealing with
$\gamma_5$.   Once the general tensor coefficients are determined, the external
particles can be treated in 4-dimensions (the 't Hooft-Veltman scheme) and
standard helicity methods used to construct the helicity amplitudes.   

By summing over helicities and colours, the full interference of tree and
two-loop graphs can be reconstructed.   In previous work~\cite{qqunlike,
qqlike},  the finite part has been separated from infrared singular parts using
the Catani formalism~\cite{catani}.  In this procedure, the singular operators
multiply tree and one-loop amplitudes.  Changing the scheme for the external
particles therefore changes the overall pole contribution.   However, the
finite remainder left after subtracting the poles is invariant under changing
from the conventional dimensional regularisation scheme to the 't Hooft-Veltman
scheme.  Therefore, the finite parts of the helicity amplitudes presented here
precisely reproduce the  finite parts of the interference of tree and two-loop
graphs  given in~\cite{qqunlike, qqlike}.   Similarly, the finite parts of the
self interference of one-loop amplitudes can be reconstructed purely from the
finite parts of the one-loop helicity amplitudes.

As in quark-gluon and gluon-gluon scattering, the presence of four coloured
external particles gives rise to additional colour correlations proportional to
$1/\epsilon$. While the existence of such additional non-trivial colour
correlations was pointed out in Ref.~\cite{catani} and are expected on general
grounds~\cite{stermanTY}, the precise form of the colour structure of  ${\bom
H}^{(2)}$ was not predicted.  So far, the results for gluon-gluon 
scattering~\cite{BFDgggg} and quark-gluon scattering~\cite{BFDqqgg,qqgghel} as
well as those for quark-quark scattering presented in this paper support the
form given in Eqs.~(\ref{eq:h2coloura}) and (\ref{eq:h2colourb}). These colour
correlations vanish when the interference of tree and two-loop amplitudes is
constructed.

As discussed in Sec.~\ref{sec:intro}, the more general goal is to produce a
numerical NNLO program (similar to the NLO codes JETRAD~\cite{jetrad1,jetrad2} and
EKS~\cite{EKS1,EKS2} that have been extensively used over the past decade) and to
reduce the theoretical uncertainty on the jet cross section to a level that
is  competitive with the anticipated experimental measurements at the Tevatron
and the LHC. Such a programme would naturally include the two-loop amplitudes
presented here. At present there is intensive
activity~\cite{Kosower:multiple,Kosower:allorders,Kosower:antenna,subsNNLO,GGHphasespace,babisphasespace}  towards achieving this goal and thereby
testing QCD at the few percent level.  

\section*{Acknowledgements}

We thank Maria Elena Tejeda-Yeomans for
help in comparing with the results of Ref.~\cite{metythesis} 
and Adrian Signer for helpful discussions.
We gratefully acknowledge the 
financial support of PPARC via a Senior Fellowship.
We thank Abilio De Freitas for pointing out several typographical errors.

\section*{Note added in proof}

After this work was completed, we became aware of a similar calculation
\cite{BFDqqQQ} that finds complete agreement with the results presented here.

\newpage
\appendix
\section{Finite two-loop contributions}
\label{app:twoloop}

In this appendix we give explicit formulae the coefficients $A,\ldots,C$ for the finite 
two-loop amplitudes defined in Section~\ref{subsec:twoloopamps} for the four quark scattering processes
of Eqs.~(\ref{eq:schannel})--(\ref{eq:tchannel2}).

As usual, the polylogarithms ${\rm Li}_n(w)$ are defined by
\begin{eqnarray}
 {\rm Li}_n(w) &=& \int_0^w \frac{dt}{t} {\rm Li}_{n-1}(t) \qquad {\rm ~for~}
 n=2,3,4\nonumber \\
 {\rm Li}_2(w) &=& -\int_0^w \frac{dt}{t} \log(1-t).
\label{eq:lidef}
\end{eqnarray} 
Using the standard polylogarithm identities~\cite{kolbig},
we retain the polylogarithms with arguments $x$, $1-x$ and
$(x-1)/x$, where
\begin{equation}
\label{eq:xydef}
x = -\frac{t}{s}, \qquad y = -\frac{u}{s} = 1-x, \qquad z=-\frac{t}{u} = \frac{x}{x-1}.
\end{equation}
For convenience, we also introduce the following logarithms
\begin{equation}
\label{eq:xydef1}
\lnx = \log\left(\frac{-t}{s}\right),
\qquad \lny = \log\left(\frac{-u}{s}\right),
\qquad \Ls = \log\left(\frac{s}{\mu^2}\right),
\end{equation}
where $\mu$ is the renormalisation scale.

\subsection{$q(p_2,+) + \bar q(p_1,-)\to Q(p_3,+)+\bar Q(p_4,-)$}

{\small
\begin{eqnarray}
{A^{(2),[1]}_{s+-+-}}&=&{\TTOUU }\,{}\Biggl ({3\over 2}\,{\Lidx}-{\Lx}\,{\Licx}+{1\over 4}\,{\Lx^2}\,{\Libx}+{\Lx}\,{\zeta_3}-{1\over 4}\,{\Lx^2}\,{\pi^2}+{1\over 16}\,{\Lx^4}+{16\over 9}\,{\Lx^2}-{1\over 60}\,{\pi^4}\xxx
-{37\over 72}\,{\Lx^3}-{29\over 36}\,{\Lx}\,{\pi^2}\Biggr ){}\xxx
+{\tou }\,{}\Biggl ({1\over 2}\,{\Licx}-{1\over 2}\,{\Lx}\,{\Libx}+{\zeta_3}-{29\over 36}\,{\pi^2}+{1\over 6}\,{\Lx^2}-{2\over 3}\,{\Lx}\,{\pi^2}+{32\over 9}\,{\Lx}+{1\over 3}\,{\Lx^3}-{1\over 4}\,{\Lx^2}\,{\Ly}\Biggr ){}\xxx
+{\one }\,{}\Biggl ({}-{1\over 2}\,{\Lidx}-{4\over 3}\,{\Licx}+{4\over 3}\,{\Lx}\,{\Libx}-{1\over 3}\,{\pi^2}\,{\Libx}+{1\over 4}\,{\Lx^2}\,{\Libx}-{7\over 2}\,{\Ly}\,{\zeta_3}+{2}\,{\Lx}\,{\zeta_3}\xxx
+{269\over 72}\,{\zeta_3}+{227\over 144}\,{\Lx}-{409\over 432}\,{\Ly}+{1\over 48}\,{\Lx^4}+{137\over 1440}\,{\pi^4}-{1\over 3}\,{\Lx}\,{\Ly}\,{\pi^2}-{37\over 72}\,{\Lx^3}-{7\over 6}\,{\Lx}\,{\pi^2}+{2\over 3}\,{\Lx^2}\,{\Ly}\xxx
+{23213\over 5184}+{1\over 6}\,{\Lx^3}\,{\Ly}+{28\over 9}\,{\Lx^2}-{25\over 18}\,{\pi^2}-{1\over 6}\,{\Lx^2}\,{\pi^2}+{11\over 48}\,{\Ly}\,{\pi^2}\Biggr ){}\xxx
+{i\pi}\,{}\Biggl ({\TTOUU }\,{}\Biggl ({}-{\Licx}+{1\over 2}\,{\Lx}\,{\Libx}+{\zeta_3}+{32\over 9}\,{\Lx}-{5\over 8}\,{\Lx^2}+{1\over 4}\,{\Lx^3}\Biggr ){}\xxx
+{\tou }\,{}\Biggl ({}-{1\over 2}\,{\Libx}+{\Lx^2}-{1\over 2}\,{\Lx}\,{\Ly}+{13\over 6}\,{\Lx}+{32\over 9}\Biggr ){}\xxx
+{\one } \,{}\Biggl ({1\over 2}\,{\Lx}\,{\Libx}+{4\over 3}\,{\Libx}-{3\over 2}\,{\zeta_3}+{55\over 9}+{4\over 3}\,{\Lx}\,{\Ly}-{1\over 6}\,{\Lx}\,{\pi^2}+{23\over 9}\,{\Lx}-{5\over 8}\,{\Lx^2}+{1\over 12}\,{\Lx^3} 
\xxx+{1\over 2}\,{\Lx^2}\,{\Ly}-{19\over 144}\,{\pi^2}\Biggr ){}\Biggr ){}\\
{B^{(2),[1]}_{s+-+-}}&=&{\tou }\,{}\Biggl ({6}\,{\Lidx}-{6}\,{\Lidy}+{6}\,{\Lidz}-{6}\,{\Ly}\,{\Licx}-{3\over 2}\,{\Licx}-{6}\,{\Licy}+{3\over 2}\,{\Lx}\,{\Libx}\xxx
+{\pi^2}\,{\Libx}+{6}\,{\Ly}\,{\Liby}+{7}\,{\zeta_3}+{6}\,{\Ly}\,{\zeta_3}-{37\over 9}\,{\Lx}+{1\over 4}\,{\Ly^4}-{\Lx}\,{\Ly^3}+{5\over 4}\,{\Lx}\,{\pi^2}+{\Lx^2}\,{\Ly}\xxx
-{7\over 60}\,{\pi^4}+{1\over 2}\,{\Ly^2}\,{\pi^2}+{17\over 12}\,{\Lx^2}+{19\over 9}\,{\pi^2}+{1\over 4}\,{\Lx}\,{\Ly}+{3}\,{\Lx}\,{\Ly^2}-{7\over 6}\,{\Ly}\,{\pi^2}-{5\over
6}\,{\Lx^3}\Biggr ){}+{5\over 2}\,{\uot }\,{\Ly^2}\xxx
+{\TTOUU }\,{}\Biggl ({11\over 2}\,{\Lidx}-{8}\,{\Lidy}+{8}\,{\Lidz}-{\Licx}-{7}\,{\Ly}\,{\Licx}+{2}\,{\Lx}\,{\Licx}+{2}\,{\Lx}\,{\Licy}\xxx
-{3\over 4}\,{\Lx^2}\,{\Libx}+{\Lx}\,{\Libx}+{2\over 3}\,{\pi^2}\,{\Libx}+{\Lx}\,{\Ly}\,{\Libx}+{7}\,{\Ly}\,{\zeta_3}+{\zeta_3}-{2}\,{\Lx}\,{\zeta_3}\xxx
-{5\over 6}\,{\Lx}\,{\Ly}\,{\pi^2}+{1\over 8}\,{\Lx^2}\,{\Ly}+{7\over 8}\,{\Lx^2}\,{\pi^2}+{1\over 3}\,{\Ly^4}-{14\over 9}\,{\Lx^2}-{1\over 6}\,{\Lx^3}\,{\Ly}\xxx
-{3\over 16}\,{\Lx^4}-{4\over 3}\,{\Lx}\,{\Ly^3}+{2\over 3}\,{\Ly^2}\,{\pi^2}+{101\over 72}\,{\Lx^3}+{\Lx^2}\,{\Ly^2}+{13\over 9}\,{\Lx}\,{\pi^2}-{1\over 60}\,{\pi^4}\Biggr ){}\xxx
+{\one } \,{}\Biggl ({}-{5\over 2}\,{\Lidx}+{2}\,{\Lidy}-{2}\,{\Lidz}+{\Lx}\,{\Licx}+{19\over 6}\,{\Licx}+{\Ly}\,{\Licx}-{2}\,{\Lx}\,{\Licy}\xxx
-{29\over 3}\,{\Licy} -{3\over 4}\,{\Lx^2}\,{\Libx}+{4\over 3}\,{\pi^2}\,{\Libx}-{19\over 6}\,{\Lx}\,{\Libx}-{2}\,{\Lx}\,{\Ly}\,{\Libx}+{29\over 3}\,{\Ly}\,{\Liby}\xxx
-{\Lx}\,{\Ly}\,{\Liby}-{3\over 2}\,{\Ly}\,{\zeta_3}-{11\over 72}\,{\zeta_3}+{5\over 2}\,{\Lx}\,{\zeta_3}-{271\over 54}\,{\Lx}+{5\over 8}\,{\Lx^2}\,{\pi^2}-{1\over
3}\,{\Lx^3}\,{\Ly}+{31\over 16}\,{\Lx}\,{\pi^2}\xxx
+{1\over 24}\,{\Ly^4}+{1\over 3}\,{\Lx}\,{\Ly}\,{\pi^2}-{275\over 54}\,{\Ly}-{403\over 72}\,{\Lx^2}+{89\over 72}\,{\Lx^3}-{1\over 16}\,{\Lx^4}-{71\over 36}\,{\Ly^3}+{779\over
72}\,{\Ly^2}+{1\over 3}\,{\Lx}\,{\Ly^3}\xxx
-{1\over 2}\,{\Ly^2}\,{\pi^2}-{41\over 24}\,{\Lx^2}\,{\Ly}+{49\over 12}\,{\Lx}\,{\Ly^2}+{5\over 2}\,{\Lx}\,{\Ly}-{7\over 4}\,{\Lx^2}\,{\Ly^2}-{475\over 144}\,{\Ly}\,{\pi^2}+{83\over
48}\,{\pi^2}-{13\over 288}\,{\pi^4}+{30659\over 1296}\Biggr ){}\xxx
+{i\pi}\,{}\Biggl ({\tou }\,{}\Biggl ({}-{6}\,{\Licx}-{9\over 2}\,{\Libx}+{6}\,{\zeta_3}-{37\over 9}+{2}\,{\Lx}\,{\Ly}-{9\over 4}\,{\Lx^2}-{1\over 12}\,{\pi^2}-{7\over 12}\,{\Lx}+{1\over
4}\,{\Ly}\Biggr ){}\xxx
+{5}\,{\uot }\,{\Ly}+{\TTOUU }\,{}\Biggl ({}-{5}\,{\Licx}+{2}\,{\Licy}-{1\over 2}\,{\Ly}\,{\Libx}-{1\over 2}\,{\Lx}\,{\Libx}+{\Libx}-{3\over 2}\,{\Ly}\,{\Liby}\xxx
+{5}\,{\zeta_3}+{1\over 12}\,{\Ly}\,{\pi^2}-{1\over 2}\,{\Lx}\,{\Ly^2}-{28\over 9}\,{\Lx}-{3\over 4}\,{\Lx^3}+{1\over 12}\,{\Lx}\,{\pi^2}-{1\over 6}\,{\pi^2}+{2}\,{\Lx^2}+{1\over
4}\,{\Lx}\,{\Ly}\Biggr ){}\xxx
+{\one } \,{}\Biggl ({2}\,{\Licx}-{2}\,{\Licy}-{77\over 6}\,{\Libx}-{5\over 2}\,{\Lx}\,{\Libx}+{1\over 2}\,{\Ly}\,{\Libx}+{3\over 2}\,{\Ly}\,{\Liby}+{\zeta_3}-{5\over
12}\,{\Ly}\,{\pi^2}\xxx
+{41\over 9}-{\Lx^2}\,{\Ly}-{3}\,{\Ly^2}-{59\over 12}\,{\Lx}\,{\Ly}-{49\over 36}\,{\Lx}+{7\over 4}\,{\Lx^2}-{1\over 4}\,{\Lx^3}+{1\over 2}\,{\Ly^3}+{25\over
36}\,{\pi^2}+{\Lx}\,{\Ly^2}\xxx
+{1\over 4}\,{\Lx}\,{\pi^2}+{473\over 36}\,{\Ly}\Biggr ){}\Biggr ){}\\
{C^{(2),[1]}_{s+-+-}}&=&{\TTOUU }\,{}\Biggl ({}-{6}\,{\Lidx}+{6}\,{\Lidy}-{6}\,{\Lidz}+{3}\,{\Licx}+{3}\,{\Ly}\,{\Licx}-{\Lx}\,{\Licx}-{6}\,{\Lx}\,{\Licy}\xxx
+{\Lx^2}\,{\Libx}-{3}\,{\Lx}\,{\Libx}+{\pi^2}\,{\Libx}-{3}\,{\Lx}\,{\Ly}\,{\Libx}-{3}\,{\Ly}\,{\zeta_3}-{3}\,{\zeta_3}+{\Lx}\,{\zeta_3}+{5\over 2}\,{\Lx}\,{\Ly}\,{\pi^2}\xxx
-{3\over 8}\,{\Lx^2}\,{\Ly}-{23\over 24}\,{\Lx^2}\,{\pi^2}-{1\over 4}\,{\Ly^4}-{9\over 2}\,{\Lx^2}+{1\over 2}\,{\Lx^3}\,{\Ly}+{1\over 6}\,{\Lx^4}+{\Lx}\,{\Ly^3}-{1\over
2}\,{\Ly^2}\,{\pi^2}-{23\over 24}\,{\Lx^3}\xxx
-{3}\,{\Lx^2}\,{\Ly^2}+{1\over 6}\,{\Lx}\,{\pi^2}-{13\over 60}\,{\pi^4}\Biggr ){}\xxx
+{\tou }\,{}\Biggl ({2}\,{\Licx}-{2}\,{\Lx}\,{\Libx}-{8}\,{\zeta_3}-{6}\,{\Lx}+{1\over 2}\,{\Ly}\,{\pi^2}-{1\over 12}\,{\Lx}\,{\pi^2}-{7\over 4}\,{\Lx^2}\,{\Ly}-{3\over
4}\,{\Lx}\,{\Ly}+{7\over 6}\,{\pi^2}\xxx
-{15\over 4}\,{\Lx^2}+{3\over 4}\,{\Lx^3}\Biggr ){}+{3\over 2}\,{\uot }\,{\Ly^2}\xxx
+{\one } \,{}\Biggl ({10}\,{\Lidx}+{2}\,{\Lidy}+{6}\,{\Lidz}-{\Licx}-{3}\,{\Ly}\,{\Licx}-{3}\,{\Lx}\,{\Licx}+{6}\,{\Lx}\,{\Licy}\xxx
-{8}\,{\Ly}\,{\Licy}+{\Lx^2}\,{\Libx}+{6}\,{\Lx}\,{\Ly}\,{\Libx}+{\Lx}\,{\Libx}+{1\over 3}\,{\pi^2}\,{\Libx}+{4}\,{\Ly^2}\,{\Liby}\xxx
+{3}\,{\Lx}\,{\Ly}\,{\Liby}+{6}\,{\Ly}\,{\zeta_3}-{4}\,{\Lx}\,{\zeta_3}-{35\over 4}\,{\zeta_3}+{93\over 16}\,{\Lx}+{7\over 24}\,{\Ly^4}-{9\over 4}\,{\Ly^3}-{7\over
8}\,{\Lx^2}\,{\pi^2}+{1\over 6}\,{\Lx^3}\,{\Ly}\xxx
+{1\over 8}\,{\Lx^2}-{13\over 24}\,{\Lx^3}+{1\over 24}\,{\Lx^4}+{2\over 3}\,{\Lx}\,{\Ly}\,{\pi^2}+{71\over 8}\,{\Ly^2}+{1\over 3}\,{\Lx}\,{\Ly^3}-{2\over 3}\,{\Ly^2}\,{\pi^2}+{7\over
8}\,{\Lx^2}\,{\Ly}+{9\over 4}\,{\Lx}\,{\Ly^2}\xxx
+{21\over 4}\,{\Lx^2}\,{\Ly^2}-{15\over 2}\,{\Lx}\,{\Ly}+{1\over 4}\,{\Ly}\,{\pi^2}-{189\over 16}\,{\Ly}+{85\over 48}\,{\pi^2}-{23\over 120}\,{\pi^4}+{511\over 64}\Biggr ){}\xxx
+{i\pi}\,{}\Biggl ({\TTOUU }\,{}\Biggl ({2}\,{\Licx}-{6}\,{\Licy}+{3\over 2}\,{\Ly}\,{\Libx}-{\Lx}\,{\Libx}-{3}\,{\Libx}+{9\over 2}\,{\Ly}\,{\Liby}-{2}\,{\zeta_3}\xxx
-{1\over 4}\,{\Ly}\,{\pi^2}+{3\over 2}\,{\Lx}\,{\Ly^2}-{9}\,{\Lx}+{2\over 3}\,{\Lx^3}-{1\over 12}\,{\Lx}\,{\pi^2}+{1\over 2}\,{\pi^2}-{7\over 4}\,{\Lx^2}-{3\over 4}\,{\Lx}\,{\Ly}\Biggr
){}\xxx
+{\tou }\,{}\Biggl ({}-{2}\,{\Libx}+{5\over 12}\,{\pi^2}-{6}+{3\over 2}\,{\Lx^2}-{7\over 2}\,{\Lx}\,{\Ly}-{3\over 4}\,{\Ly}-{33\over 4}\,{\Lx}\Biggr ){}+{3}\,{\uot }\,{\Ly}+{\one } \nonumber \\ 
 &&\,{}\Biggl ({}-{6}\,{\Licx}-{2}\,{\Licy}+{\Libx}+{5}\,{\Lx}\,{\Libx}+{9\over 2}\,{\Ly}\,{\Libx}+{19\over 2}\,{\Ly}\,{\Liby}+{2}\,{\zeta_3}-{3\over 4}\,{\Ly}\,{\pi^2}\xxx
 -{6}+{1\over 2}\,{\Lx^2}\,{\Ly}-{9\over 2}\,{\Ly^2}+{25\over 4}\,{\Lx}\,{\Ly}-{29\over 4}\,{\Lx}-{5\over 4}\,{\Lx^2}+{1\over 6}\,{\Lx^3}+{1\over 6}\,{\Ly^3}\xxx
 -{1\over 12}\,{\pi^2}+{7}\,{\Lx}\,{\Ly^2}+{1\over 12}\,{\Lx}\,{\pi^2}+{41\over 4}\,{\Ly}\Biggr ){}\Biggr ){}\\
{D^{(2),[1]}_{s+-+-}}&=&{\tou }\,{}\Biggl ({}-{8\over 9}\,{\Lx}+{2\over 9}\,{\pi^2}-{1\over 6}\,{\Lx^2}\Biggr ){}+{\TTOUU }\,{}\Biggl ({1\over 18}\,{\Lx^3}-{4\over 9}\,{\Lx^2}+{2\over
9}\,{\Lx}\,{\pi^2}\Biggr ){}\xxx
+{\one } \,{}\Biggl ({1\over 3}\,{\Licx}-{1\over 3}\,{\Lx}\,{\Libx}-{49\over 36}\,{\zeta_3}+{1\over 36}\,{\Lx}-{1\over 6}\,{\Lx^2}\,{\Ly}-{19\over 36}\,{\Lx^2}+{1\over
18}\,{\Lx^3}-{455\over 108}\xxx
+{19\over 24}\,{\pi^2}-{1\over 24}\,{\Ly}\,{\pi^2}+{1\over 3}\,{\Lx}\,{\pi^2}+{25\over 108}\,{\Ly}\Biggr ){}\xxx
+{i\pi}\,{}\Biggl ({\tou }\,{}\Biggl ({}-{8\over 9}-{2\over 3}\,{\Lx}\Biggr ){}-{8\over 9}\,{\TTOUU }\,{\Lx}\xxx
+{\one } \,{}\Biggl ({}-{1\over 3}\,{\Libx}-{361\over 108}-{1\over 3}\,{\Lx}\,{\Ly}+{5\over 72}\,{\pi^2}-{7\over 18}\,{\Lx}\Biggr ){}\Biggr ){}\\
{E^{(2),[1]}_{s+-+-}}&=&{\TTOUU }\,{}\Biggl ({8\over 9}\,{\Lx^2}-{4\over 9}\,{\Lx}\,{\pi^2}-{1\over 9}\,{\Lx^3}\Biggr ){}+{\tou }\,{}\Biggl ({1\over 3}\,{\Lx^2}+{16\over 9}\,{\Lx}-{4\over
9}\,{\pi^2}\Biggr ){}\xxx
+{\one } \,{}\Biggl ({}-{2\over 3}\,{\Licx}+{2\over 3}\,{\Licy}+{2\over 3}\,{\Lx}\,{\Libx}-{2\over 3}\,{\Ly}\,{\Liby}-{35\over 36}\,{\zeta_3}+{2\over 9}\,{\Ly^3}-{1\over
3}\,{\Lx}\,{\Ly^2}-{685\over 162}\xxx
+{77\over 72}\,{\Ly}\,{\pi^2}-{1\over 9}\,{\Lx^3}+{223\over 108}\,{\Ly}-{13\over 24}\,{\pi^2}-{5\over 8}\,{\Lx}\,{\pi^2}+{1\over 3}\,{\Lx^2}\,{\Ly}-{31\over 108}\,{\Lx}+{19\over
18}\,{\Lx^2}-{29\over 18}\,{\Ly^2}\Biggr ){}\xxx
+{i\pi}\,{}\Biggl ({16\over 9}\,{\TTOUU }\,{\Lx}+{\tou }\,{}\Biggl ({4\over 3}\,{\Lx}+{16\over 9}\Biggr ){}+{\one } \,{}\Biggl ({4\over 3}\,{\Libx}-{23\over 36}-{11\over 9}\,{\Ly}+{7\over 9}\,{\Lx}+{2\over 3}\,{\Lx}\,{\Ly}-{1\over 9}\,{\pi^2}\Biggr ){}\Biggr ){}\\
{F^{(2),[1]}_{s+-+-}}&=&{\one } \,{}\Biggl ({25\over 81}-{1\over 9}\,{\pi^2}\Biggr ){}+{10\over 27}\,{i\pi}\,{\one } \nonumber \\ 
 &&\\
{A^{(2),[2]}_{s+-+-}}&=&{\TTOUU }\,{}\Biggl ({}-{5}\,{\Lidx}+{8}\,{\Lidy}-{8}\,{\Lidz}-{1\over 2}\,{\Licx}+{7}\,{\Ly}\,{\Licx}-{2}\,{\Lx}\,{\Licx}-{2}\,{\Lx}\,{\Licy}\xxx
+{1\over 2}\,{\Lx^2}\,{\Libx}+{1\over 2}\,{\Lx}\,{\Libx}-{1\over 3}\,{\pi^2}\,{\Libx}-{\Lx}\,{\Ly}\,{\Libx}-{7}\,{\Ly}\,{\zeta_3}+{1\over 2}\,{\zeta_3}+{2}\,{\Lx}\,{\zeta_3}\xxx
+{5\over 6}\,{\Lx}\,{\Ly}\,{\pi^2}-{1\over 8}\,{\Lx^2}\,{\Ly}+{2\over 3}\,{\Lx}\,{\pi^2}-{1\over 3}\,{\Ly^4}-{5\over 4}\,{\Lx^2}+{1\over 6}\,{\Lx^3}\,{\Ly}-{2\over
3}\,{\Ly^2}\,{\pi^2}-{3\over 8}\,{\Lx^2}\,{\pi^2}\xxx
-{\Lx^2}\,{\Ly^2}+{4\over 3}\,{\Lx}\,{\Ly^3}-{2\over 45}\,{\pi^4}\Biggr ){}\xxx
+{\tou }\,{}\Biggl ({}-{6}\,{\Lidx}+{6}\,{\Lidy}-{6}\,{\Lidz}+{6}\,{\Ly}\,{\Licx}+{2}\,{\Licx}+{6}\,{\Licy}-{2}\,{\Lx}\,{\Libx}\xxx
-{\pi^2}\,{\Libx}-{6}\,{\Ly}\,{\Liby}-{6}\,{\Ly}\,{\zeta_3}-{7}\,{\zeta_3}+{1\over 3}\,{\pi^2}+{\Lx}\,{\Ly^3}-{1\over 4}\,{\Ly^4}+{7\over 60}\,{\pi^4}-{1\over 2}\,{\Ly^2}\,{\pi^2}\xxx
-{1\over 12}\,{\Lx}\,{\pi^2}-{3\over 4}\,{\Lx^2}\,{\Ly}-{7\over 4}\,{\Lx}\,{\Ly}-{3}\,{\Lx}\,{\Ly^2}+{5\over 6}\,{\Ly}\,{\pi^2}-{\Lx^2}\Biggr ){}-{3\over 2}\,{\uot }\,{\Ly^2}\xxx
+{\one}\,{}\Biggl ({\Lidx}-{3}\,{\Lidy}-{2}\,{\Lidz}-{2}\,{\Lx}\,{\Licx}+{2}\,{\Ly}\,{\Licx}+{\Licx}+{28\over 3}\,{\Licy}+{3}\,{\Ly}\,{\Licy}\xxx
-{\Lx}\,{\Licy}-{\Lx}\,{\Libx}+{1\over 2}\,{\Lx^2}\,{\Libx}-{\Lx}\,{\Ly}\,{\Libx}+{\pi^2}\,{\Libx}-{1\over 2}\,{\Ly^2}\,{\Liby}-{28\over 3}\,{\Ly}\,{\Liby}\xxx
-{5\over 2}\,{\Lx}\,{\zeta_3}+{2}\,{\Ly}\,{\zeta_3}-{737\over 72}\,{\zeta_3}+{13\over 6}\,{\Lx}\,{\Ly}\,{\pi^2}-{409\over 432}\,{\Lx}-{1\over 4}\,{\Lx^2}\,{\pi^2}-{1\over
8}\,{\Ly^4}+{11\over 18}\,{\Ly^3}+{1673\over 432}\,{\Ly}\xxx
-{37\over 48}\,{\Lx}\,{\pi^2}-{385\over 72}\,{\Ly^2}+{1\over 8}\,{\Lx^2}\,{\Ly}-{5\over 8}\,{\Ly^2}\,{\pi^2}-{23213\over 5184}-{47\over 12}\,{\Lx}\,{\Ly^2}+{139\over
72}\,{\Ly}\,{\pi^2}-{13\over 8}\,{\Lx^2}\,{\Ly^2}\xxx
+{5\over 12}\,{\Lx^3}\,{\Ly}-{19\over 8}\,{\Lx}\,{\Ly}+{11\over 12}\,{\Lx}\,{\Ly^3}+{2\over 3}\,{\pi^2}-{161\over 1440}\,{\pi^4}\Biggr ){}\xxx
+{i\pi}\,{}\Biggl ({\TTOUU }\,{}\Biggl ({5}\,{\Licx}-{2}\,{\Licy}+{1\over 2}\,{\Ly}\,{\Libx}+{1\over 2}\,{\Libx}+{3\over 2}\,{\Ly}\,{\Liby}-{5}\,{\zeta_3}+{1\over 12}\,{\Lx}\,{\pi^2}\xxx
-{5\over 2}\,{\Lx}-{1\over 12}\,{\Ly}\,{\pi^2}-{1\over 4}\,{\Lx}\,{\Ly}+{1\over 2}\,{\Lx}\,{\Ly^2}-{1\over 12}\,{\pi^2}-{3\over 8}\,{\Lx^2}+{1\over 6}\,{\Lx^3}\Biggr ){}\xxx
+{\tou }\,{}\Biggl ({6}\,{\Licx}+{4}\,{\Libx}-{6}\,{\zeta_3}-{3\over 2}\,{\Lx}\,{\Ly}+{1\over 4}\,{\Lx^2}+{1\over 4}\,{\pi^2}-{15\over 4}\,{\Lx}-{7\over 4}\,{\Ly}\Biggr ){}-{3}\,{\uot
}\,{\Ly}\xxx
+{\one}\,{}\Biggl ({2}\,{\Licy}+{25\over 3}\,{\Libx}+{1\over 2}\,{\Ly}\,{\Libx}+{1\over 2}\,{\Ly}\,{\Liby}-{1\over 2}\,{\zeta_3}-{23\over 9}+{1\over 2}\,{\Lx}\,{\pi^2}\xxx
-{1\over 2}\,{\Ly}\,{\pi^2}+{1\over 2}\,{\Lx}\,{\Ly^2}+{3\over 4}\,{\Ly^2}+{1\over 4}\,{\Ly^3}-{19\over 8}\,{\Lx}+{5\over 8}\,{\Lx^2}+{5\over 12}\,{\Lx^3}-{\Lx^2}\,{\Ly}-{13\over
144}\,{\pi^2}\xxx
+{7\over 4}\,{\Lx}\,{\Ly}-{545\over 72}\,{\Ly}\Biggr ){}\Biggr ){}\\
{B^{(2),[2]}_{s+-+-}}&=&{\TTOUU }\,{}\Biggl ({3\over 2}\,{\Lidx}-{4}\,{\Lidy}+{4}\,{\Lidz}-{1\over 2}\,{\Licx}-{2}\,{\Ly}\,{\Licx}+{2}\,{\Lx}\,{\Licx}+{4}\,{\Lx}\,{\Licy}\xxx
-{3\over 4}\,{\Lx^2}\,{\Libx}+{1\over 2}\,{\Lx}\,{\Libx}-{\pi^2}\,{\Libx}+{2}\,{\Lx}\,{\Ly}\,{\Libx}+{2}\,{\Ly}\,{\zeta_3}+{1\over 2}\,{\zeta_3}-{2}\,{\Lx}\,{\zeta_3}\xxx
-{5\over 3}\,{\Lx}\,{\Ly}\,{\pi^2}+{1\over 4}\,{\Lx^2}\,{\Ly}+{1\over 3}\,{\Lx^2}\,{\pi^2}+{1\over 6}\,{\Ly^4}+{127\over 36}\,{\Lx^2}-{1\over 3}\,{\Lx^3}\,{\Ly}+{1\over 48}\,{\Lx^4}-{2\over
3}\,{\Lx}\,{\Ly^3}\xxx
+{1\over 3}\,{\Ly^2}\,{\pi^2}-{11\over 36}\,{\Lx^3}+{2}\,{\Lx^2}\,{\Ly^2}-{47\over 36}\,{\Lx}\,{\pi^2}+{41\over 180}\,{\pi^4}\Biggr ){}\xxx
+{\tou }\,{}\Biggl ({}-{5\over 2}\,{\Licx}+{5\over 2}\,{\Lx}\,{\Libx}+{4}\,{\zeta_3}-{83\over 36}\,{\pi^2}+{5\over 4}\,{\Lx^2}\,{\Ly}+{32\over 9}\,{\Lx}-{1\over 3}\,{\Lx}\,{\pi^2}+{17\over
12}\,{\Lx^2}\xxx
-{1\over 12}\,{\Lx^3}+{2}\,{\Lx}\,{\Ly}\Biggr ){}-{2}\,{\uot }\,{\Ly^2}\xxx
+{\one}\,{}\Biggl ({}-{9\over 2}\,{\Lidx}-{\Ly}\,{\Licx}-{7\over 3}\,{\Licx}+{3}\,{\Lx}\,{\Licx}+{1\over 3}\,{\Licy}+{2}\,{\Ly}\,{\Licy}-{\Lx}\,{\Licy}\xxx
-{2}\,{\pi^2}\,{\Libx}-{3\over 4}\,{\Lx^2}\,{\Libx}-{\Lx}\,{\Ly}\,{\Libx}+{7\over 3}\,{\Lx}\,{\Libx}-{2}\,{\Ly^2}\,{\Liby}-{\Lx}\,{\Ly}\,{\Liby}\xxx
-{1\over 3}\,{\Ly}\,{\Liby}-{\Lx}\,{\zeta_3}+{695\over 72}\,{\zeta_3}+{\Ly}\,{\zeta_3}+{1\over 6}\,{\Lx}\,{\pi^2}+{39\over 8}\,{\Lx}\,{\Ly}-{5\over 2}\,{\Lx}\,{\Ly}\,{\pi^2}+{5\over
12}\,{\Lx^2}\,{\Ly}+{103\over 36}\,{\Ly^3}\xxx
+{25\over 18}\,{\Lx}+{3\over 8}\,{\Lx^2}\,{\pi^2}+{29\over 36}\,{\Ly}\,{\pi^2}-{30659\over 1296}-{1\over 8}\,{\Ly^4}-{193\over 18}\,{\Ly^2}+{1\over 48}\,{\Lx^4}+{13\over
8}\,{\Ly^2}\,{\pi^2}-{5\over 12}\,{\Lx^3}\,{\Ly}\xxx
-{5\over 3}\,{\Lx}\,{\Ly^2}-{17\over 12}\,{\Lx}\,{\Ly^3}-{1\over 8}\,{\Lx^2}\,{\Ly^2}+{49\over 6}\,{\Ly}+{125\over 72}\,{\Lx^2}-{7\over 18}\,{\Lx^3}-{241\over 144}\,{\pi^2}+{209\over
1440}\,{\pi^4}\Biggr ){}\xxx
+{i\pi}\,{}\Biggl ({\TTOUU }\,{}\Biggl ({4}\,{\Licy}+{1\over 2}\,{\Libx}+{1\over 2}\,{\Lx}\,{\Libx}-{\Ly}\,{\Libx}-{3}\,{\Ly}\,{\Liby}+{1\over 6}\,{\Ly}\,{\pi^2}+{1\over 2}\,{\Lx}\,{\Ly}\xxx
-{\Lx}\,{\Ly^2}+{127\over 18}\,{\Lx}-{1\over 12}\,{\Lx^3}-{1\over 12}\,{\pi^2}-{1\over 6}\,{\Lx}\,{\pi^2}\Biggr ){}\xxx
+{\tou }\,{}\Biggl ({5\over 2}\,{\Libx}+{2}\,{\Ly}+{20\over 3}\,{\Lx}+{32\over 9}+{5\over 2}\,{\Lx}\,{\Ly}-{1\over 2}\,{\pi^2}-{1\over 4}\,{\Lx^2}\Biggr ){}-{4}\,{\uot }\,{\Ly}\xxx
+{\one}\,{}\Biggl ({2}\,{\Licx}+{\Licy}+{8\over 3}\,{\Libx}-{4}\,{\Ly}\,{\Libx}-{3\over 2}\,{\Lx}\,{\Libx}-{8}\,{\Ly}\,{\Liby}+{\Lx^2}\,{\Ly}\xxx
+{17\over 12}\,{\Ly}\,{\pi^2}-{46\over 9}-{7\over 12}\,{\Lx}\,{\pi^2}+{21\over 4}\,{\Ly^2}-{11\over 12}\,{\Ly^3}+{337\over 72}\,{\Lx}-{\Lx^2}-{1\over 3}\,{\Lx^3}-{17\over 36}\,{\pi^2}-{13\over
6}\,{\Lx}\,{\Ly}\xxx
-{6}\,{\Lx}\,{\Ly^2}-{797\over 72}\,{\Ly}\Biggr ){}\Biggr ){}\\
{C^{(2),[2]}_{s+-+-}}&=&{\TTOUU }\,{}\Biggl ({5\over 2}\,{\Lidx}-{2}\,{\Lidy}+{2}\,{\Lidz}-{\Licx}-{\Ly}\,{\Licx}+{2}\,{\Lx}\,{\Licy}-{1\over 4}\,{\Lx^2}\,{\Libx}\xxx
+{\Lx}\,{\Ly}\,{\Libx}+{\Lx}\,{\Libx}-{1\over 3}\,{\pi^2}\,{\Libx}+{\zeta_3}+{\Ly}\,{\zeta_3}+{1\over 12}\,{\Ly^4}+{1\over 8}\,{\Lx^2}\,{\Ly}-{1\over 6}\,{\Lx^3}\,{\Ly}\xxx
-{5\over 6}\,{\Lx}\,{\Ly}\,{\pi^2}+{2}\,{\Lx^2}+{3\over 8}\,{\Lx^3}-{1\over 16}\,{\Lx^4}-{1\over 3}\,{\Lx}\,{\Ly^3}+{1\over 6}\,{\Ly^2}\,{\pi^2}+{3\over 8}\,{\Lx^2}\,{\pi^2}+{\Lx^2}\,{\Ly^2}\xxx
-{1\over 6}\,{\Lx}\,{\pi^2}+{1\over 15}\,{\pi^4}\Biggr ){}\xxx
+{\one}\,{}\Biggl ({}-{7\over 2}\,{\Lidx}-{\Lidy}-{2}\,{\Lidz}+{1\over 2}\,{\Licx}+{\Lx}\,{\Licx}+{\Ly}\,{\Licx}+{3}\,{\Ly}\,{\Licy}\xxx
-{2}\,{\Lx}\,{\Licy}-{1\over 3}\,{\pi^2}\,{\Libx}-{1\over 4}\,{\Lx^2}\,{\Libx}-{1\over 2}\,{\Lx}\,{\Libx}-{2}\,{\Lx}\,{\Ly}\,{\Libx}-{3\over 2}\,{\Ly^2}\,{\Liby}\xxx
-{\Lx}\,{\Ly}\,{\Liby}-{4}\,{\Ly}\,{\zeta_3}+{3}\,{\Lx}\,{\zeta_3}+{23\over 4}\,{\zeta_3}+{7\over 24}\,{\Lx^2}\,{\pi^2}-{1\over 6}\,{\Lx}\,{\pi^2}-{45\over 16}\,{\Lx}+{5\over 2}\,{\Lx}\,{\Ly}\xxx
-{1\over 3}\,{\Lx}\,{\Ly}\,{\pi^2}+{3\over 4}\,{\Ly^3}+{5\over 24}\,{\Lx^3}-{1\over 48}\,{\Lx^4}-{511\over 64}-{1\over 12}\,{\Ly^4}-{29\over 8}\,{\Ly^2}-{1\over 6}\,{\Lx}\,{\Ly^3}+{1\over
6}\,{\Ly^2}\,{\pi^2}\xxx
-{3\over 8}\,{\Lx^2}\,{\Ly}-{3\over 4}\,{\Lx}\,{\Ly^2}-{7\over 4}\,{\Lx^2}\,{\Ly^2}+{1\over 12}\,{\Ly}\,{\pi^2}+{93\over 16}\,{\Ly}+{5\over 8}\,{\Lx^2}-{53\over 48}\,{\pi^2}+{13\over
120}\,{\pi^4}\Biggr ){}\xxx
+{\tou }\,{}\Biggl ({}-{1\over 2}\,{\Licx}+{1\over 2}\,{\Lx}\,{\Libx}+{3}\,{\zeta_3}+{7\over 4}\,{\Lx^2}-{1\over 6}\,{\Ly}\,{\pi^2}-{1\over 2}\,{\pi^2}+{1\over 4}\,{\Lx}\,{\Ly}+{3}\,{\Lx}-{1\over
6}\,{\Lx^3}\xxx
-{1\over 12}\,{\Lx}\,{\pi^2}+{1\over 2}\,{\Lx^2}\,{\Ly}\Biggr ){}-{1\over 2}\,{\uot }\,{\Ly^2}\xxx
+{i\pi}\,{}\Biggl ({\TTOUU }\,{}\Biggl ({}-{\Licx}+{2}\,{\Licy}-{1\over 2}\,{\Ly}\,{\Libx}+{1\over 2}\,{\Lx}\,{\Libx}+{\Libx}-{3\over 2}\,{\Ly}\,{\Liby}+{\zeta_3}\xxx
+{1\over 12}\,{\Ly}\,{\pi^2}-{1\over 2}\,{\Lx}\,{\Ly^2}+{4}\,{\Lx}-{1\over 4}\,{\Lx^3}+{1\over 12}\,{\Lx}\,{\pi^2}-{1\over 6}\,{\pi^2}+{3\over 4}\,{\Lx^2}+{1\over 4}\,{\Lx}\,{\Ly}\Biggr ){}\xxx
+{\one}\,{}\Biggl ({2}\,{\Licx}+{\Licy}-{1\over 2}\,{\Libx}-{3\over 2}\,{\Lx}\,{\Libx}-{3\over 2}\,{\Ly}\,{\Libx}-{7\over 2}\,{\Ly}\,{\Liby}-{\zeta_3}+{1\over 4}\,{\Ly}\,{\pi^2}\xxx
-{1\over 12}\,{\Lx}\,{\pi^2}+{3\over 2}\,{\Ly^2}+{15\over 4}\,{\Lx}+{1\over 2}\,{\Lx^2}-{1\over 12}\,{\Lx^3}+{3}+{1\over 12}\,{\pi^2}-{9\over 4}\,{\Lx}\,{\Ly}-{5\over 2}\,{\Lx}\,{\Ly^2}-{19\over
4}\,{\Ly}\Biggr ){}\xxx
+{\tou }\,{}\Biggl ({1\over 2}\,{\Libx}-{1\over 12}\,{\pi^2}+{1\over 4}\,{\Ly}+{15\over 4}\,{\Lx}-{1\over 4}\,{\Lx^2}+{3}+{\Lx}\,{\Ly}\Biggr ){}-{\uot }\,{\Ly}\Biggr ){}\\
{D^{(2),[2]}_{s+-+-}}&=&{\one}\,{}\Biggl ({}-{1\over 3}\,{\Licy}+{1\over 3}\,{\Ly}\,{\Liby}+{49\over 36}\,{\zeta_3}+{29\over 36}\,{\Ly^2}-{149\over 108}\,{\Ly}-{1\over
24}\,{\Lx}\,{\pi^2}+{455\over 108}+{1\over 6}\,{\Lx}\,{\Ly^2}\xxx
-{17\over 36}\,{\Ly}\,{\pi^2}-{41\over 72}\,{\pi^2}-{1\over 9}\,{\Ly^3}+{25\over 108}\,{\Lx}\Biggr ){}\xxx
+{i\pi}\,{\one}\,{}\Biggl ({}-{1\over 3}\,{\Libx}+{11\over 18}\,{\Ly}-{1\over 72}\,{\pi^2}+{265\over 108}\Biggr ){}\\
{E^{(2),[2]}_{s+-+-}}&=&{\TTOUU }\,{}\Biggl ({1\over 18}\,{\Lx^3}-{4\over 9}\,{\Lx^2}+{2\over 9}\,{\Lx}\,{\pi^2}\Biggr ){}+{\tou }\,{}\Biggl ({}-{8\over 9}\,{\Lx}+{2\over 9}\,{\pi^2}-{1\over
6}\,{\Lx^2}\Biggr ){}\xxx
+{\one}\,{}\Biggl ({1\over 3}\,{\Licx}-{1\over 3}\,{\Licy}-{1\over 3}\,{\Lx}\,{\Libx}+{1\over 3}\,{\Ly}\,{\Liby}+{35\over 36}\,{\zeta_3}+{685\over 162}+{29\over 36}\,{\Ly^2}+{1\over
3}\,{\Lx}\,{\pi^2}\xxx
-{5\over 9}\,{\Ly}\,{\pi^2}-{19\over 36}\,{\Lx^2}-{1\over 9}\,{\Ly^3}-{11\over 12}\,{\Ly}+{23\over 72}\,{\pi^2}+{1\over 18}\,{\Lx^3}+{1\over 36}\,{\Lx}-{1\over 6}\,{\Lx^2}\,{\Ly}+{1\over
6}\,{\Lx}\,{\Ly^2}\Biggr ){}\xxx
+{i\pi}\,{}\Biggl ({}-{8\over 9}\,{\TTOUU }\,{\Lx}+{\tou }\,{}\Biggl ({}-{8\over 9}-{2\over 3}\,{\Lx}\Biggr ){}\xxx
+{\one}\,{}\Biggl ({}-{2\over 3}\,{\Libx}+{1\over 18}\,{\pi^2}-{1\over
3}\,{\Lx}\,{\Ly}+{11\over 18}\,{\Ly}+{55\over 36}-{7\over 18}\,{\Lx}\Biggr ){}\Biggr ){}\xxx\\
{F^{(2),[2]}_{s+-+-}}&=&{\one}\,{}\Biggl ({}-{25\over 81}+{1\over 9}\,{\pi^2}\Biggr ){}-{10\over 27}\,{i\pi}\,{\one} 
\end{eqnarray}
}
\subsection{$q(p_2,+) + \bar q(p_1,-)\to Q(p_3,-)+\bar Q(p_4,+)$}
{\small
\begin{eqnarray}
{A^{(2),[1]}_{s+--+}}&=&{3\over 2}\,{\tou }\,{\Lx^2}+{\UUOTT }\,{}\Biggl ({}-{3}\,{\Lidx}+{3}\,{\Lidy}-{3}\,{\Lidz}-{3}\,{\Lx}\,{\Licy}-{1\over 2}\,{\pi^2}\,{\Libx}+{3}\,{\Lx}\,{\zeta_3}\xxx
+{1\over 2}\,{\Lx}\,{\Ly^3}-{1\over 30}\,{\pi^4}-{3\over 4}\,{\Lx^2}\,{\Ly^2}-{1\over 4}\,{\Ly^2}\,{\pi^2}-{1\over 8}\,{\Ly^4}\Biggr ){}\xxx
+{\uot }\,{}\Biggl ({}-{3}\,{\Lidx}+{3}\,{\Lidy}-{3}\,{\Lidz}-{3}\,{\Licx}-{3}\,{\Lx}\,{\Licy}+{3}\,{\Lx}\,{\Libx}-{1\over 2}\,{\pi^2}\,{\Libx}\xxx+{3}\,{\Lx}\,{\zeta_3}
+{3}\,{\zeta_3}-{1\over 8}\,{\Ly^4}-{3\over 4}\,{\Lx^2}\,{\Ly^2}+{3\over 2}\,{\Lx^2}\,{\Ly}-{1\over 2}\,{\Lx}\,{\pi^2}+{1\over 2}\,{\pi^2}+{1\over 2}\,{\Lx}\,{\Ly^3}-{1\over 30}\,{\pi^4}-{1\over
4}\,{\Ly^2}\,{\pi^2}\Biggr ){}\xxx
+{\one}\,{}\Biggl ({\Lidx}-{29\over 6}\,{\Licx}-{\Lx}\,{\Licx}+{29\over 6}\,{\Lx}\,{\Libx}+{1\over 2}\,{\Lx^2}\,{\Libx}-{1\over 3}\,{\pi^2}\,{\Libx}-{7\over 2}\,{\Ly}\,{\zeta_3}\xxx
+{3}\,{\Lx}\,{\zeta_3}+{413\over 72}\,{\zeta_3}-{95\over 48}\,{\Lx}-{409\over 432}\,{\Ly}+{1\over 12}\,{\Lx^4}+{113\over 1440}\,{\pi^4}-{49\over 36}\,{\Lx^3}-{65\over 36}\,{\Lx}\,{\pi^2}+{29\over
12}\,{\Lx^2}\,{\Ly}\xxx
+{23213\over 5184}+{1\over 6}\,{\Lx^3}\,{\Ly}+{56\over 9}\,{\Lx^2}-{1\over 12}\,{\pi^2}-{5\over 12}\,{\Lx^2}\,{\pi^2}-{1\over 3}\,{\Lx}\,{\Ly}\,{\pi^2}+{11\over 48}\,{\Ly}\,{\pi^2}\Biggr ){}\xxx
+{i\pi}\,{}\Biggl ({3}\,{\tou }\,{\Lx}+{\UUOTT }\,{}\Biggl ({}-{3}\,{\Licy}+{3}\,{\zeta_3}\Biggr ){}+{\uot }\,{}\Biggl ({}-{3}\,{\Licy}+{3}\,{\Libx}+{3}\,{\zeta_3}+{3}\,{\Lx}\,{\Ly}-{1\over
2}\,{\pi^2}\Biggr ){}\xxx
+{\one}\,{}\Biggl ({}-{\Licx}+{\Lx}\,{\Libx}+{29\over 6}\,{\Libx}-{1\over 2}\,{\zeta_3}+{23\over 9}-{1\over 6}\,{\Lx}\,{\pi^2}+{125\over 18}\,{\Lx}+{29\over 6}\,{\Lx}\,{\Ly}\xxx
+{1\over 3}\,{\Lx^3}+{1\over 2}\,{\Lx^2}\,{\Ly}-{91\over 144}\,{\pi^2}-{9\over 4}\,{\Lx^2}\Biggr ){}\Biggr ){}\\
{B^{(2),[1]}_{s+--+}}&=&{\UUOTT }\,{}\Biggl ({4}\,{\Lidx}-{2}\,{\Lidy}+{4}\,{\Lidz}+{2}\,{\Ly}\,{\Licx}-{\Ly}\,{\Licy}-{\Licy}+{5}\,{\Lx}\,{\Licy}\xxx
+{4\over 3}\,{\pi^2}\,{\Libx}-{3\over 2}\,{\Lx}\,{\Ly}\,{\Libx}+{\Ly}\,{\Liby}-{1\over 2}\,{\Lx}\,{\Ly}\,{\Liby}-{5}\,{\Lx}\,{\zeta_3}+{\zeta_3}+{\Ly}\,{\zeta_3}+{73\over 18}\,{\Ly^2}\xxx
-{5\over 6}\,{\Lx}\,{\Ly^3}+{1\over 4}\,{\Ly^4}+{1\over 24}\,{\Ly^2}\,{\pi^2}-{59\over 72}\,{\Ly^3}+{1\over 2}\,{\Lx^2}\,{\Ly^2}+{1\over 45}\,{\pi^4}+{1\over 8}\,{\Lx}\,{\Ly^2}-{13\over
9}\,{\Ly}\,{\pi^2}+{1\over 12}\,{\Lx}\,{\Ly}\,{\pi^2}\Biggr ){}\xxx
-{7\over 2}\,{\tou }\,{\Lx^2}+{\uot }\,{}\Biggl ({6}\,{\Lidx}-{6}\,{\Lidy}+{6}\,{\Lidz}+{6}\,{\Licx}+{6}\,{\Lx}\,{\Licy}+{\pi^2}\,{\Libx}\xxx
-{6}\,{\Lx}\,{\Libx}-{2}\,{\zeta_3}-{6}\,{\Lx}\,{\zeta_3}+{3\over 2}\,{\Lx^2}\,{\Ly^2}+{1\over 4}\,{\Ly^4}+{1\over 15}\,{\pi^4}+{7\over 12}\,{\Ly^2}-{\Lx}\,{\Ly^3}+{64\over 9}\,{\Ly}+{5\over
6}\,{\Lx}\,{\pi^2}\xxx
-{3}\,{\Lx^2}\,{\Ly}+{1\over 2}\,{\Ly^2}\,{\pi^2}+{1\over 4}\,{\Ly^3}-{25\over 9}\,{\pi^2}+{1\over 4}\,{\Lx}\,{\Ly}+{1\over 4}\,{\Lx}\,{\Ly^2}-{13\over 12}\,{\Ly}\,{\pi^2}\Biggr ){}\xxx
+{\one}\,{}\Biggl ({}-{\Lidx}-{4}\,{\Lidy}+{2}\,{\Lidz}+{3}\,{\Lx}\,{\Licx}+{29\over 3}\,{\Licx}-{2}\,{\Ly}\,{\Licx}+{\Lx}\,{\Licy}\xxx
-{8\over 3}\,{\Licy}+{\Ly}\,{\Licy}-{3\over 2}\,{\Lx^2}\,{\Libx}+{2\over 3}\,{\pi^2}\,{\Libx}-{29\over 3}\,{\Lx}\,{\Libx}+{1\over 2}\,{\Lx}\,{\Ly}\,{\Libx}\xxx
-{1\over 2}\,{\Lx}\,{\Ly}\,{\Liby}+{8\over 3}\,{\Ly}\,{\Liby}-{3\over 2}\,{\Ly}\,{\zeta_3}-{659\over 72}\,{\zeta_3}-{1\over 2}\,{\Lx}\,{\zeta_3}+{3\over 2}\,{\Lx^2}\,{\pi^2}-{49\over
54}\,{\Lx}+{427\over 144}\,{\Lx}\,{\pi^2}\xxx
+{1\over 8}\,{\Ly^4}-{7\over 12}\,{\Lx}\,{\Ly}\,{\pi^2}-{1\over 2}\,{\Lx^3}\,{\Ly}-{869\over 72}\,{\Lx^2}+{125\over 36}\,{\Lx^3}-{1\over 4}\,{\Lx^4}-{65\over 72}\,{\Ly^3}+{349\over
72}\,{\Ly^2}-{1\over 6}\,{\Lx}\,{\Ly^3}\xxx
+{1\over 8}\,{\Ly^2}\,{\pi^2}-{67\over 12}\,{\Lx^2}\,{\Ly}+{29\over 24}\,{\Lx}\,{\Ly^2}+{5\over 2}\,{\Lx}\,{\Ly}+{1\over 4}\,{\Lx^2}\,{\Ly^2}-{85\over 48}\,{\Ly}\,{\pi^2}+{109\over
54}\,{\Ly}-{455\over 144}\,{\pi^2}\xxx
+{143\over 1440}\,{\pi^4}+{30659\over 1296}\Biggr ){}\xxx
+{i\pi}\,{}\Biggl ({\UUOTT }\,{}\Biggl ({2}\,{\Licx}+{4}\,{\Licy}-{\Libx}+{1\over 2}\,{\Ly}\,{\Libx}-{\Lx}\,{\Libx}+{3\over 2}\,{\Ly}\,{\Liby}-{4}\,{\zeta_3}\xxx
-{1\over 6}\,{\Ly}\,{\pi^2}+{1\over 3}\,{\Ly^3}-{3\over 4}\,{\Lx}\,{\Ly}+{1\over 2}\,{\Lx}\,{\Ly^2}-{\Ly^2}+{73\over 9}\,{\Ly}\Biggr ){}-{7}\,{\tou }\,{\Lx}\xxx
+{\uot }\,{}\Biggl ({6}\,{\Licy}-{6}\,{\Libx}-{6}\,{\zeta_3}+{64\over 9}-{11\over 2}\,{\Lx}\,{\Ly}+{1\over 4}\,{\Lx}+{3\over 4}\,{\pi^2}+{\Ly^2}+{61\over 12}\,{\Ly}\Biggr ){}\xxx
+{\one}\,{}\Biggl ({\Licx}+{2}\,{\Licy}-{37\over 3}\,{\Libx}-{2}\,{\Lx}\,{\Libx}-{1\over 2}\,{\Ly}\,{\Libx}-{3\over 2}\,{\Ly}\,{\Liby}-{2}\,{\zeta_3}-{1\over 6}\,{\Ly}\,{\pi^2}\xxx
+{142\over 9}-{\Ly^2}+{1\over 6}\,{\Ly^3}-{383\over 36}\,{\Lx}+{6}\,{\Lx^2}-{\Lx^3}-{\Lx^2}\,{\Ly}+{49\over 36}\,{\pi^2}-{137\over 12}\,{\Lx}\,{\Ly}+{1\over 3}\,{\Lx}\,{\pi^2}+{175\over
36}\,{\Ly}\Biggr ){}\Biggr ){}\xxx\\
{C^{(2),[1]}_{s+--+}}&=&{\UUOTT }\,{}\Biggl ({6}\,{\Lidx}+{6}\,{\Lidz}-{6}\,{\Ly}\,{\Licx}+{3}\,{\Lx}\,{\Licy}-{5}\,{\Ly}\,{\Licy}+{3}\,{\Licy}+{9\over 2}\,{\Lx}\,{\Ly}\,{\Libx}\xxx
-{\pi^2}\,{\Libx}+{2}\,{\Ly^2}\,{\Liby}+{3\over 2}\,{\Lx}\,{\Ly}\,{\Liby}-{3}\,{\Ly}\,{\Liby}+{5}\,{\Ly}\,{\zeta_3}-{3}\,{\zeta_3}-{3}\,{\Lx}\,{\zeta_3}+{3}\,{\Lx^2}\,{\Ly^2}\xxx
+{3\over 2}\,{\Ly^2}+{1\over 3}\,{\Ly^4}+{5\over 24}\,{\Ly^2}\,{\pi^2}-{1\over 2}\,{\Lx}\,{\Ly^3}-{7\over 24}\,{\Ly^3}-{3\over 8}\,{\Lx}\,{\Ly^2}-{7\over 6}\,{\Ly}\,{\pi^2}-{1\over
4}\,{\Lx}\,{\Ly}\,{\pi^2}\Biggr ){}\xxx
+{\uot }\,{}\Biggl ({4}\,{\Licy}-{4}\,{\Ly}\,{\Liby}-{4}\,{\zeta_3}-{17\over 12}\,{\Ly}\,{\pi^2}+{1\over 2}\,{\Lx}\,{\pi^2}+{6}\,{\Ly}-{11\over 4}\,{\Lx}\,{\Ly^2}-{3\over 4}\,{\Lx}\,{\Ly}\xxx
+{7\over 4}\,{\Ly^3}+{9\over 4}\,{\Ly^2}-{1\over 6}\,{\pi^2}\Biggr ){}+{3\over 2}\,{\tou }\,{\Lx^2}\xxx
+{\one}\,{}\Biggl ({}-{2}\,{\Lidx}+{8}\,{\Lidy}-{6}\,{\Lidz}+{6}\,{\Ly}\,{\Licx}-{4}\,{\Lx}\,{\Licx}-{3}\,{\Lx}\,{\Licy}-{3}\,{\Ly}\,{\Licy}\xxx
+{\Licy}-{3\over 2}\,{\Lx}\,{\Ly}\,{\Libx}+{2}\,{\Lx^2}\,{\Libx}+{7\over 3}\,{\pi^2}\,{\Libx}-{\Ly}\,{\Liby}+{2}\,{\Ly^2}\,{\Liby}+{3\over 2}\,{\Lx}\,{\Ly}\,{\Liby}\xxx
-{2}\,{\Ly}\,{\zeta_3}-{19\over 4}\,{\zeta_3}-{5\over 24}\,{\Ly^3}-{7\over 24}\,{\Ly^4}+{189\over 16}\,{\Lx}-{15\over 2}\,{\Lx}\,{\Ly}+{2\over 3}\,{\Lx^3}\,{\Ly}+{41\over
12}\,{\Lx}\,{\Ly}\,{\pi^2}+{3\over 4}\,{\Lx}\,{\pi^2}\xxx
+{5\over 24}\,{\Lx^4}-{11\over 6}\,{\Lx^2}\,{\pi^2}+{65\over 8}\,{\Ly^2}-{9\over 4}\,{\Lx^3}-{11\over 8}\,{\Ly^2}\,{\pi^2}+{9\over 4}\,{\Lx^2}\,{\Ly}-{1\over 8}\,{\Lx}\,{\Ly^2}+{11\over
6}\,{\Lx}\,{\Ly^3}\xxx
-{3\over 4}\,{\Lx^2}\,{\Ly^2}-{1\over 2}\,{\Ly}\,{\pi^2}-{93\over 16}\,{\Ly}+{7\over 8}\,{\Lx^2}+{7\over 16}\,{\pi^2}-{49\over 120}\,{\pi^4}+{511\over 64}\Biggr ){}\xxx
+{i\pi}\,{}\Biggl ({\UUOTT }\,{}\Biggl ({}-{6}\,{\Licx}-{2}\,{\Licy}+{3}\,{\Libx}+{3\over 2}\,{\Ly}\,{\Libx}+{3}\,{\Lx}\,{\Libx}+{5\over 2}\,{\Ly}\,{\Liby}\xxx
+{2}\,{\zeta_3}+{1\over 3}\,{\Ly}\,{\pi^2}+{1\over 3}\,{\Ly^3}+{9\over 4}\,{\Lx}\,{\Ly}+{3\over 2}\,{\Lx}\,{\Ly^2}+{1\over 4}\,{\Ly^2}+{3}\,{\Ly}\Biggr ){}\xxx
+{\uot }\,{}\Biggl ({4}\,{\Libx}+{5\over 12}\,{\pi^2}-{3\over 4}\,{\Lx}-{3\over 2}\,{\Lx}\,{\Ly}+{9\over 2}\,{\Ly^2}+{15\over 4}\,{\Ly}+{6}\Biggr ){}+{3}\,{\tou }\,{\Lx}\xxx
+{\one}\,{}\Biggl ({2}\,{\Licx}-{6}\,{\Licy}+{\Libx}+{\Lx}\,{\Libx}+{9\over 2}\,{\Ly}\,{\Libx}+{23\over 2}\,{\Ly}\,{\Liby}-{2}\,{\zeta_3}-{4\over 3}\,{\Ly}\,{\pi^2}\xxx
+{6}-{1\over 4}\,{\Ly^2}-{1\over 6}\,{\Ly^3}-{23\over 4}\,{\Lx}-{9\over 2}\,{\Lx^2}+{5\over 6}\,{\Lx^3}+{1\over 2}\,{\Lx^2}\,{\Ly}+{5\over 12}\,{\pi^2}+{21\over
4}\,{\Lx}\,{\Ly}\xxx
+{7}\,{\Lx}\,{\Ly^2}+{35\over 4}\,{\Ly}\Biggr ){}\Biggr ){}\xxx\\
{D^{(2),[1]}_{s+--+}}&=&{\one}\,{}\Biggl ({1\over 3}\,{\Licx}-{1\over 3}\,{\Lx}\,{\Libx}-{49\over 36}\,{\zeta_3}+{5\over 9}\,{\Lx}\,{\pi^2}-{1\over 6}\,{\Lx^2}\,{\Ly}+{1\over 9}\,{\Lx^3}+{11\over
12}\,{\Lx}-{455\over 108}-{29\over 36}\,{\Lx^2}\xxx
-{1\over 24}\,{\Ly}\,{\pi^2}+{25\over 108}\,{\Ly}+{41\over 72}\,{\pi^2}\Biggr ){}+{i\pi}\,{\one}\,{}\Biggl ({}-{1\over 3}\,{\Libx}-{265\over 108}-{1\over 3}\,{\Lx}\,{\Ly}+{5\over 72}\,{\pi^2}-{11\over 18}\,{\Lx}\Biggr ){}\\
{E^{(2),[1]}_{s+--+}}&=&{\UUOTT }\,{}\Biggl ({}-{8\over 9}\,{\Ly^2}+{4\over 9}\,{\Ly}\,{\pi^2}+{1\over 9}\,{\Ly^3}\Biggr ){}+{\uot }\,{}\Biggl ({4\over 9}\,{\pi^2}-{1\over 3}\,{\Ly^2}-{16\over
9}\,{\Ly}\Biggr ){}\xxx
+{\one}\,{}\Biggl ({}-{2\over 3}\,{\Licx}+{2\over 3}\,{\Licy}+{2\over 3}\,{\Lx}\,{\Libx}-{2\over 3}\,{\Ly}\,{\Liby}-{35\over 36}\,{\zeta_3}+{29\over 18}\,{\Lx^2}+{5\over 8}\,{\Ly}\,{\pi^2}\xxx
-{77\over 72}\,{\Lx}\,{\pi^2}+{1\over 3}\,{\Lx^2}\,{\Ly}-{19\over 18}\,{\Ly^2}-{2\over 9}\,{\Lx^3}-{223\over 108}\,{\Lx}+{25\over 72}\,{\pi^2}-{685\over 162}-{1\over 3}\,{\Lx}\,{\Ly^2}+{31\over
108}\,{\Ly}+{1\over 9}\,{\Ly^3}\Biggr ){}\xxx
+{i\pi}\,{}\Biggl ({}-{16\over 9}\,{\UUOTT }\,{\Ly}+{\uot }\,{}\Biggl
({}-{16\over 9}-{4\over 3}\,{\Ly}\Biggr ){}\xxx
+{\one}\,{}\Biggl ({4\over 3}\,{\Libx}-{151\over 36}+{2\over
3}\,{\Lx}\,{\Ly}+{11\over 9}\,{\Lx}-{1\over 9}\,{\pi^2}-{7\over 9}\,{\Ly}\Biggr ){}\Biggr ){}\xxx\\
{F^{(2),[1]}_{s+--+}}&=&{\one}\,{}\Biggl ({25\over 81}-{1\over 9}\,{\pi^2}\Biggr ){}+{10\over 27}\,{i\pi}\,{\one}\\
{A^{(2),[2]}_{s+--+}}&=&{\UUOTT }\,{}\Biggl ({}-{3}\,{\Lidx}-{1\over 2}\,{\Lidy}-{3}\,{\Lidz}+{2}\,{\Ly}\,{\Licy}+{3\over 2}\,{\Licy}-{3}\,{\Lx}\,{\Licy}-{1\over 6}\,{\pi^2}\,{\Libx}\xxx
-{1\over 4}\,{\Ly^2}\,{\Liby}-{3\over 2}\,{\Ly}\,{\Liby}-{2}\,{\Ly}\,{\zeta_3}+{3}\,{\Lx}\,{\zeta_3}-{3\over 2}\,{\zeta_3}-{3\over 4}\,{\Lx^2}\,{\Ly^2}+{1\over 3}\,{\Lx}\,{\Ly}\,{\pi^2}-{7\over
48}\,{\Ly^4}\xxx
+{1\over 180}\,{\pi^4}-{37\over 36}\,{\Ly^2}+{1\over 2}\,{\Lx}\,{\Ly^3}+{11\over 36}\,{\Ly^3}-{1\over 36}\,{\Ly}\,{\pi^2}+{1\over 6}\,{\Ly^2}\,{\pi^2}\Biggr ){}+{1\over 2}\,{\tou }\,{\Lx^2}\xxx
+{\uot }\,{}\Biggl ({}-{3}\,{\Lidx}+{3}\,{\Lidy}-{3}\,{\Lidz}-{3}\,{\Licx}-{3\over 2}\,{\Licy}-{3}\,{\Lx}\,{\Licy}-{1\over 2}\,{\pi^2}\,{\Libx}\xxx
+{3}\,{\Lx}\,{\Libx}+{3\over 2}\,{\Ly}\,{\Liby}+{\zeta_3}+{3}\,{\Lx}\,{\zeta_3}+{1\over 12}\,{\Ly^3}+{3\over 2}\,{\Lx^2}\,{\Ly}-{1\over 8}\,{\Ly^4}-{1\over 6}\,{\Lx}\,{\pi^2}-{32\over
9}\,{\Ly}\xxx
+{7\over 12}\,{\Ly^2}+{1\over 2}\,{\Lx}\,{\Ly^3}+{1\over 2}\,{\Ly}\,{\pi^2}+{1\over 4}\,{\Lx}\,{\Ly^2}+{3\over 2}\,{\Lx}\,{\Ly}-{3\over 4}\,{\Lx^2}\,{\Ly^2}-{1\over 4}\,{\Ly^2}\,{\pi^2}+{5\over
36}\,{\pi^2}-{1\over 30}\,{\pi^4}\Biggr ){}\xxx
+{\one}\,{}\Biggl ({2}\,{\Lidx}+{5\over 2}\,{\Lidy}-{4}\,{\Lidz}+{3}\,{\Ly}\,{\Licx}-{9\over 2}\,{\Licx}-{4}\,{\Lx}\,{\Licx}+{\Ly}\,{\Licy}\xxx
+{1\over 3}\,{\Licy}-{3}\,{\Lx}\,{\Licy}+{9\over 2}\,{\Lx}\,{\Libx}+{4\over 3}\,{\pi^2}\,{\Libx}-{2}\,{\Lx}\,{\Ly}\,{\Libx}+{\Lx^2}\,{\Libx}\xxx
-{1\over 4}\,{\Ly^2}\,{\Liby}-{1\over 3}\,{\Ly}\,{\Liby}+{3}\,{\Ly}\,{\zeta_3}-{1\over 2}\,{\Lx}\,{\zeta_3}-{17\over 72}\,{\zeta_3}-{409\over 432}\,{\Lx}+{7\over 12}\,{\Lx^3}\,{\Ly}-{5\over
8}\,{\Lx^2}\,{\pi^2}\xxx
-{3\over 16}\,{\Ly^4}+{7\over 18}\,{\Ly^3}+{137\over 432}\,{\Ly}+{3\over 4}\,{\Lx^2}-{23213\over 5184}+{8\over 3}\,{\Lx}\,{\Ly}\,{\pi^2}+{5\over 4}\,{\Lx}\,{\Ly^3}-{29\over
24}\,{\Ly^2}\,{\pi^2}-{3\over 16}\,{\Lx}\,{\pi^2}\xxx
-{125\over 72}\,{\Ly^2}+{9\over 4}\,{\Lx^2}\,{\Ly}-{21\over 8}\,{\Lx^2}\,{\Ly^2}-{1\over 8}\,{\Lx}\,{\Ly}-{2\over 3}\,{\Lx}\,{\Ly^2}+{13\over 8}\,{\Ly}\,{\pi^2}+{35\over 36}\,{\pi^2}-{449\over
1440}\,{\pi^4}\Biggr ){}\xxx
+{i\pi}\,{}\Biggl ({\UUOTT }\,{}\Biggl ({}-{\Licy}+{3\over 2}\,{\Libx}-{1\over 2}\,{\Ly}\,{\Liby}+{\zeta_3}+{3\over 4}\,{\Ly^2}-{1\over 4}\,{\Ly^3}+{3\over 2}\,{\Lx}\,{\Ly}-{37\over
18}\,{\Ly}\Biggr ){}\xxx
+{\tou }\,{\Lx}+{\uot }\,{}\Biggl ({}-{3}\,{\Licy}+{3\over 2}\,{\Libx}+{3}\,{\zeta_3}-{32\over 9}+{2}\,{\Lx}\,{\Ly}+{3\over 2}\,{\Lx}-{1\over 4}\,{\pi^2}-{1\over 4}\,{\Ly^2}+{5\over
6}\,{\Ly}\Biggr ){}\xxx
+{\one}\,{}\Biggl ({}-{\Licx}-{2}\,{\Licy}+{\Ly}\,{\Libx}+{29\over 6}\,{\Libx}+{5\over 2}\,{\Ly}\,{\Liby}+{5\over 2}\,{\zeta_3}-{\Lx^2}\,{\Ly}+{7\over 12}\,{\Lx}\,{\pi^2}\xxx
-{7\over 12}\,{\Ly}\,{\pi^2}+{\Lx}\,{\Ly^2}-{1\over 4}\,{\Ly^2}+{1\over 2}\,{\Ly^3}+{7\over 12}\,{\Lx^3}-{55\over 9}-{97\over 144}\,{\pi^2}+{11\over 8}\,{\Lx}+{7\over 2}\,{\Lx}\,{\Ly}+{5\over 72}\,{\Ly}\Biggr ){}\Biggr ){}\\
{B^{(2),[2]}_{s+--+}}&=&{\one}\,{}\Biggl ({}-{\Lidx}-{4}\,{\Lidy}+{6}\,{\Lidz}+{5}\,{\Lx}\,{\Licx}-{1\over 3}\,{\Licx}-{5}\,{\Ly}\,{\Licx}+{4}\,{\Lx}\,{\Licy}\xxx
+{11\over 6}\,{\Licy}-{3\over 2}\,{\Lx^2}\,{\Libx}+{1\over 3}\,{\Lx}\,{\Libx}-{3}\,{\pi^2}\,{\Libx}+{5\over 2}\,{\Lx}\,{\Ly}\,{\Libx}-{11\over 6}\,{\Ly}\,{\Liby}\xxx
-{1\over 2}\,{\Lx}\,{\Ly}\,{\Liby}-{\Ly^2}\,{\Liby}-{4}\,{\Lx}\,{\zeta_3}+{335\over 72}\,{\zeta_3}+{5}\,{\Ly}\,{\zeta_3}+{17\over 24}\,{\Lx^2}\,{\pi^2}
-{47\over 12}\,{\Lx}\,{\Ly}\,{\pi^2}\xxx+{49\over 72}\,{\Ly^3}-{47\over 36}\,{\Lx}\,{\pi^2}+{1\over 4}\,{\Ly^4}-{7\over 12}\,{\Lx^2}\,{\Ly}+{21\over 8}\,{\Lx}\,{\Ly}-{11\over 18}\,{\Lx^3}+{1\over
12}\,{\Ly}\,{\pi^2}-{30659\over 1296}-{13\over 6}\,{\Lx}\xxx
-{283\over 36}\,{\Ly^2}+{1\over 24}\,{\Lx^4}+{2}\,{\Ly^2}\,{\pi^2}-{3\over 4}\,{\Lx^3}\,{\Ly}-{13\over 24}\,{\Lx}\,{\Ly^2}-{9\over 4}\,{\Lx}\,{\Ly^3}+{23\over 8}\,{\Lx^2}\,{\Ly^2}+{29\over
18}\,{\Ly}\xxx
+{277\over 72}\,{\Lx^2}+{109\over 48}\,{\pi^2}+{179\over 480}\,{\pi^4}\Biggr ){}\xxx
+{\UUOTT }\,{}\Biggl ({}-{2}\,{\Lidx}-{2}\,{\Lidz}+{2}\,{\Ly}\,{\Licx}+{2}\,{\Ly}\,{\Licy}-{\Lx}\,{\Licy}-{5\over 2}\,{\Licy}-{3\over 2}\,{\Lx}\,{\Ly}\,{\Libx}\xxx
-{\Ly^2}\,{\Liby}-{1\over 2}\,{\Lx}\,{\Ly}\,{\Liby}+{5\over 2}\,{\Ly}\,{\Liby}-{2}\,{\Ly}\,{\zeta_3}+{5\over 2}\,{\zeta_3}+{\Lx}\,{\zeta_3}-{\Lx^2}\,{\Ly^2}-{127\over 36}\,{\Ly^2}\xxx
+{55\over 72}\,{\Ly^3}-{5\over 24}\,{\Ly^4}-{1\over 24}\,{\Ly^2}\,{\pi^2}-{1\over 4}\,{\Lx}\,{\Ly}\,{\pi^2}+{1\over 8}\,{\Lx}\,{\Ly^2}+{77\over 36}\,{\Ly}\,{\pi^2}+{1\over
6}\,{\Lx}\,{\Ly^3}\Biggr ){}+{1\over 2}\,{\tou }\,{\Lx^2}\xxx
+{\uot }\,{}\Biggl ({}-{\Licy}+{\Ly}\,{\Liby}+{\zeta_3}+{5\over 4}\,{\Lx}\,{\Ly^2}-{5\over 4}\,{\Lx}\,{\Ly}-{59\over 9}\,{\Ly}-{13\over 6}\,{\Ly^2}+{17\over 12}\,{\Ly}\,{\pi^2}\xxx
-{17\over 12}\,{\Ly^3}+{77\over 36}\,{\pi^2}-{1\over 2}\,{\Lx}\,{\pi^2}\Biggr ){}\xxx
+{i\pi}\,{}\Biggl ({\one}\,{}\Biggl ({4}\,{\Licy}+{13\over 6}\,{\Libx}-{7\over 2}\,{\Ly}\,{\Libx}-{17\over 2}\,{\Ly}\,{\Liby}+{\zeta_3}-{137\over 9}-{3\over 4}\,{\Lx}\,{\pi^2}+{19\over
12}\,{\Ly}\,{\pi^2}\xxx
-{11\over 2}\,{\Lx}\,{\Ly^2}+{3\over 2}\,{\Ly^2}-{7\over 12}\,{\Ly^3}+{347\over 72}\,{\Lx}-{3\over 4}\,{\Lx^2}-{5\over 12}\,{\Lx^3}+{\Lx^2}\,{\Ly}-{11\over 36}\,{\pi^2}-{5\over
12}\,{\Lx}\,{\Ly}-{679\over 72}\,{\Ly}\Biggr ){}\xxx
+{\UUOTT }\,{}\Biggl ({2}\,{\Licx}+{\Licy}-{5\over 2}\,{\Libx}-{\Lx}\,{\Libx}-{3\over 2}\,{\Ly}\,{\Libx}-{5\over 2}\,{\Ly}\,{\Liby}-{\zeta_3}-{3\over 2}\,{\Lx}\,{\Ly^2}\xxx
-{1\over 3}\,{\Ly^3}-{9\over 4}\,{\Lx}\,{\Ly}+{1\over 4}\,{\Ly^2}-{127\over 18}\,{\Ly}\Biggr ){}+{\tou }\,{\Lx}\xxx
+{\uot }\,{}\Biggl ({}-{\Libx}+{3\over 2}\,{\Lx}\,{\Ly}-{1\over 4}\,{\pi^2}-{59\over 9}-{89\over 12}\,{\Ly}-{5\over 4}\,{\Lx}-{7\over 2}\,{\Ly^2}\Biggr ){}\Biggr ){}\\
{C^{(2),[2]}_{s+--+}}&=&{\UUOTT }\,{}\Biggl ({}-{2}\,{\Lidx}-{1\over 2}\,{\Lidy}-{2}\,{\Lidz}+{2}\,{\Ly}\,{\Licx}-{\Licy}-{\Lx}\,{\Licy}+{2}\,{\Ly}\,{\Licy}\xxx
+{1\over 3}\,{\pi^2}\,{\Libx}-{3\over 2}\,{\Lx}\,{\Ly}\,{\Libx}+{\Ly}\,{\Liby}-{1\over 2}\,{\Lx}\,{\Ly}\,{\Liby}-{3\over 4}\,{\Ly^2}\,{\Liby}+{\zeta_3}+{\Lx}\,{\zeta_3}\xxx
-{2}\,{\Ly}\,{\zeta_3}+{1\over 2}\,{\Ly}\,{\pi^2}-{1\over 8}\,{\Ly^2}\,{\pi^2}+{1\over 24}\,{\Ly^3}-{\Ly^2}+{1\over 12}\,{\Lx}\,{\Ly}\,{\pi^2}-{\Lx^2}\,{\Ly^2}-{5\over 48}\,{\Ly^4}+{1\over
8}\,{\Lx}\,{\Ly^2}\xxx
+{1\over 6}\,{\Lx}\,{\Ly^3}+{1\over 180}\,{\pi^4}\Biggr ){}-{1\over 2}\,{\tou }\,{\Lx^2}\xxx
+{\uot }\,{}\Biggl ({}-{3\over 2}\,{\Licy}+{3\over 2}\,{\Ly}\,{\Liby}+{\zeta_3}-{1\over 6}\,{\Lx}\,{\pi^2}-{2\over 3}\,{\Ly^3}-{3}\,{\Ly}+{7\over 12}\,{\Ly}\,{\pi^2}-{5\over 4}\,{\Ly^2}\xxx
+{1\over 4}\,{\Lx}\,{\Ly}+{1\over 6}\,{\pi^2}+{\Lx}\,{\Ly^2}\Biggr ){}\xxx
+{\one}\,{}\Biggl ({\Lidx}-{5\over 2}\,{\Lidy}+{2}\,{\Lidz}+{\Lx}\,{\Licx}-{2}\,{\Ly}\,{\Licx}-{1\over 2}\,{\Licy}+{\Lx}\,{\Licy}+{\Ly}\,{\Licy}\xxx
-{\pi^2}\,{\Libx}-{1\over 2}\,{\Lx^2}\,{\Libx}+{1\over 2}\,{\Lx}\,{\Ly}\,{\Libx}+{1\over 2}\,{\Ly}\,{\Liby}-{1\over 2}\,{\Lx}\,{\Ly}\,{\Liby}-{3\over 4}\,{\Ly^2}\,{\Liby}\xxx
+{15\over 4}\,{\zeta_3}+{2}\,{\Lx}\,{\zeta_3}-{\Ly}\,{\zeta_3}+{2\over 3}\,{\Lx^2}\,{\pi^2}-{3\over 4}\,{\Lx^2}\,{\Ly}-{5\over 12}\,{\Lx}\,{\pi^2}+{1\over 3}\,{\Ly}\,{\pi^2}-{5\over
4}\,{\Lx}\,{\Ly}\,{\pi^2}+{1\over 24}\,{\Ly^3}\xxx
-{93\over 16}\,{\Lx}-{1\over 12}\,{\Lx^4}-{511\over 64}+{5\over 48}\,{\Ly^4}-{27\over 8}\,{\Ly^2}-{2\over 3}\,{\Lx}\,{\Ly^3}+{11\over 24}\,{\Ly^2}\,{\pi^2}-{1\over 6}\,{\Lx^3}\,{\Ly}+{1\over
8}\,{\Lx}\,{\Ly^2}\xxx
+{5\over 2}\,{\Lx}\,{\Ly}+{1\over 4}\,{\Lx^2}\,{\Ly^2}+{45\over 16}\,{\Ly}+{3\over 8}\,{\Lx^2}+{3\over 4}\,{\Lx^3}-{7\over 16}\,{\pi^2}+{61\over 360}\,{\pi^4}\Biggr ){}\xxx
+{i\pi}\,{}\Biggl ({\UUOTT }\,{}\Biggl ({2}\,{\Licx}+{\Licy}-{\Libx}-{1\over 2}\,{\Ly}\,{\Libx}-{\Lx}\,{\Libx}-{\Ly}\,{\Liby}-{\zeta_3}-{1\over 6}\,{\Ly}\,{\pi^2}\xxx
-{1\over 12}\,{\Ly^3}-{3\over 4}\,{\Lx}\,{\Ly}-{1\over 2}\,{\Lx}\,{\Ly^2}-{1\over 4}\,{\Ly^2}-{2}\,{\Ly}\Biggr ){}-{\tou }\,{\Lx}\xxx
+{\uot }\,{}\Biggl ({}-{3\over 2}\,{\Libx}+{1\over 2}\,{\Lx}\,{\Ly}-{1\over 6}\,{\pi^2}-{3}-{7\over 4}\,{\Ly^2}-{9\over 4}\,{\Ly}+{1\over 4}\,{\Lx}\Biggr ){}\xxx
+{\one}\,{}\Biggl ({}-{\Licx}+{2}\,{\Licy}-{1\over 2}\,{\Libx}-{3\over 2}\,{\Ly}\,{\Libx}-{4}\,{\Ly}\,{\Liby}+{\zeta_3}+{1\over 2}\,{\Ly}\,{\pi^2}-{5\over 2}\,{\Lx}\,{\Ly^2}\xxx
+{13\over 4}\,{\Lx}+{3\over 2}\,{\Lx^2}-{7\over 4}\,{\Lx}\,{\Ly}+{1\over 12}\,{\Ly^3}-{1\over 6}\,{\pi^2}-{3}-{1\over 3}\,{\Lx^3}-{17\over 4}\,{\Ly}\Biggr ){}\Biggr ){}\\
{D^{(2),[2]}_{s+--+}}&=&{\UUOTT }\,{}\Biggl ({}-{2\over 9}\,{\Ly}\,{\pi^2}-{1\over 18}\,{\Ly^3}+{4\over 9}\,{\Ly^2}\Biggr ){}+{\one}\,{}\Biggl ({}-{1\over 3}\,{\Licy}+{1\over
3}\,{\Ly}\,{\Liby}+{49\over 36}\,{\zeta_3}-{1\over 24}\,{\Lx}\,{\pi^2}\xxx
-{19\over 24}\,{\pi^2}+{19\over 36}\,{\Ly^2}-{1\over 4}\,{\Ly}\,{\pi^2}+{455\over 108}+{1\over 6}\,{\Lx}\,{\Ly^2}+{25\over 108}\,{\Lx}-{1\over 18}\,{\Ly^3}-{53\over 108}\,{\Ly}\Biggr ){}\xxx
+{\uot }\,{}\Biggl ({1\over 6}\,{\Ly^2}-{2\over 9}\,{\pi^2}+{8\over 9}\,{\Ly}\Biggr ){}\xxx
+{i\pi}\,{}\Biggl ({8\over 9}\,{\UUOTT }\,{\Ly}+{\one}\,{}\Biggl ({}-{1\over 3}\,{\Libx}+{361\over 108}+{7\over 18}\,{\Ly}-{1\over 72}\,{\pi^2}\Biggr ){}+{\uot }\,{}\Biggl ({2\over 3}\,{\Ly}+{8\over 9}\Biggr ){}\Biggr ){}\\
{E^{(2),[2]}_{s+--+}}&=&{\one}\,{}\Biggl ({1\over 3}\,{\Licx}-{1\over 3}\,{\Licy}-{1\over 3}\,{\Lx}\,{\Libx}+{1\over 3}\,{\Ly}\,{\Liby}+{35\over 36}\,{\zeta_3}-{1\over 8}\,{\pi^2}+{11\over
12}\,{\Lx}+{5\over 9}\,{\Lx}\,{\pi^2}\xxx
-{1\over 18}\,{\Ly^3}-{1\over 6}\,{\Lx^2}\,{\Ly}+{1\over 9}\,{\Lx^3}+{19\over 36}\,{\Ly^2}-{1\over 36}\,{\Ly}+{685\over 162}+{1\over 6}\,{\Lx}\,{\Ly^2}-{29\over 36}\,{\Lx^2}-{1\over
3}\,{\Ly}\,{\pi^2}\Biggr ){}\xxx
+{\UUOTT }\,{}\Biggl ({}-{2\over 9}\,{\Ly}\,{\pi^2}-{1\over 18}\,{\Ly^3}+{4\over 9}\,{\Ly^2}\Biggr ){}+{\uot }\,{}\Biggl ({1\over 6}\,{\Ly^2}-{2\over 9}\,{\pi^2}+{8\over 9}\,{\Ly}\Biggr ){}\xxx
+{i\pi}\,{}\Biggl ({\one}\,{}\Biggl ({}-{2\over 3}\,{\Libx}+{119\over 36}+{7\over 18}\,{\Ly}+{1\over 18}\,{\pi^2}-{1\over 3}\,{\Lx}\,{\Ly}-{11\over 18}\,{\Lx}\Biggr ){}+{8\over
9}\,{\UUOTT}\,{\Ly}\xxx
+{\uot }\,{}\Biggl ({2\over 3}\,{\Ly}+{8\over 9}\Biggr ){}\Biggr ){}\xxx\\
{F^{(2),[2]}_{s+--+}}&=&{\one}\,{}\Biggl ({}-{25\over 81}+{1\over 9}\,{\pi^2}\Biggr ){}-{10\over 27}\,{i\pi}\,{\one} 
\end{eqnarray}
}

\subsection{$q(p_2,+) + \bar Q(p_1,-)\to q(p_3,+)+\bar Q(p_4,-)$}

{\small

}

\subsection{$q(p_2,+) + \bar Q(p_1,+)\to q(p_3,+)+\bar Q(p_4,+)$}
{\small
%
}

\subsection{$q(p_2,+) +  Q(p_1,+)\to Q(p_3,+)+ q(p_4,+)$}
{\small
%
}

\subsection{$q(p_2,+) +  Q(p_1,-)\to Q(p_3,-)+ q(p_4,+)$}
{\small
%
}

\subsection{$q(p_2,+) +  Q(p_1,+)\to q(p_3,+)+ Q(p_4,+)$}
The coefficients for this amplitude are obtained by exchanging $t$ and $u$ in the
coefficients for $q(p_2,+) +  Q(p_1,+)\to Q(p_3,+)+ q(p_4,+)$,
\begin{eqnarray}
{A^{(2),[1]}_{t^\prime ++++}}&=&{A^{(2),[1]}_{u ++++}} (t \leftrightarrow u)\\
{A^{(2),[2]}_{t^\prime ++++}}&=&{A^{(2),[2]}_{u ++++}} (t \leftrightarrow u) 
\end{eqnarray}
etc.
\subsection{$q(p_2,+) +  Q(p_1,-)\to q(p_3,+)+ Q(p_4,-)$}
The coefficients for this amplitude are obtained by exchanging $t$ and $u$ in the
coefficients for $q(p_2,+) +  Q(p_1,-)\to Q(p_3,-)+ q(p_4,+)$,
\begin{eqnarray}
{A^{(2),[1]}_{t^\prime +-+-}}&=&{A^{(2),[1]}_{u +--+}} (t \leftrightarrow u)\\
{A^{(2),[2]}_{t^\prime +-+-}}&=&{A^{(2),[2]}_{u +--+}} (t \leftrightarrow u) 
\end{eqnarray}
etc.

\newpage
\section{Finite one-loop contributions}
\label{app:oneloop}

In this appendix we list the coefficients $A,\ldots,C$ for the finite one-loop amplitudes defined in Section~\ref{subsec:oneloopamps} for the four quark scattering processes
of Eqs.~(\ref{eq:schannel})--(\ref{eq:tchannel2}).
.
\subsection{$q(p_2,+) + \bar q(p_1,-)\to Q(p_3,+)+\bar Q(p_4,-)$}
\begin{eqnarray}
{A^{(1),[1]}_{s+-+-}}&=&{1\over 2}\,{\tou }\,{\Lx}+{1\over 4}\,{\TTOUU }\,{\Lx^2}+{\one}\,{}\Biggl ({1\over 4}\,{\Lx^2}-{\Lx}+{13\over 18}\Biggr ){}\xxx+{i\pi}\,{}\Biggl ({\one}\,{}\Biggl ({5\over 6}+{1\over 2}\,{\Lx}\Biggr ){}+{1\over 2}\,{\TTOUU }\,{\Lx}+{1\over 2}\,{\tou }\Biggr ){}\\
{B^{(1),[1]}_{s+-+-}}&=&{}-{\tou }\,{\Lx}-{1\over 2}\,{\TTOUU }\,{\Lx^2}+{\one}\,{}\Biggl ({2}\,{\Lx}+{\Ly^2}-{1\over 2}\,{\Lx^2}-{3}\,{\Ly}+{4}\Biggr ){}\xxx+{i\pi}\,{}\Biggl ({}-{\TTOUU }\,{\Lx}-{\tou }+{\one}\,{}\Biggl ({}-{\Lx}+{2}\,{\Ly}-{1}\Biggr ){}\Biggr ){}\\
{C^{(1),[1]}_{s+-+-}}&=&{}-{1\over 3}\,{i\pi}\,{\one}-{5\over 9}\,{\one}\\
{A^{(1),[2]}_{s+-+-}}&=&{\one}\,{}\Biggl ({}-{13\over 18}-{1\over 2}\,{\Ly^2}+{3\over 2}\,{\Ly}\Biggr ){}\xxx+{i\pi}\,{\one}\,{}\Biggl ({}-{\Ly}-{1\over 3}\Biggr ){}\\
{B^{(1),[2]}_{s+-+-}}&=&{\one}\,{}\Biggl ({}-{\Lx}+{1\over 4}\,{\Lx^2}+{3\over 2}\,{\Ly}-{1\over 2}\,{\Ly^2}-{4}\Biggr ){}+{1\over 4}\,{\TTOUU }\,{\Lx^2}+{1\over 2}\,{\tou }\,{\Lx}\xxx+{i\pi}\,{}\Biggl ({1\over 2}\,{\TTOUU }\,{\Lx}+{1\over 2}\,{\tou }+{\one}\,{}\Biggl ({1\over 2}+{1\over 2}\,{\Lx}-{\Ly}\Biggr ){}\Biggr ){}\\
{C^{(1),[2]}_{s+-+-}}&=&{5\over 9}\,{\one}+{1\over 3}\,{i\pi}\,{\one}
\end{eqnarray}

\subsection{$q(p_2,+) + \bar q(p_1,-)\to Q(p_3,-)+\bar Q(p_4,+)$}
\begin{eqnarray}
{A^{(1),[1]}_{s+--+}}&=&{\one}\,{}\Biggl ({1\over 2}\,{\Lx^2}-{3\over 2}\,{\Lx}+{13\over 18}\Biggr ){}\xxx+{i\pi}\,{\one}\,{}\Biggl ({1\over 3}+{\Lx}\Biggr ){}\\
{B^{(1),[1]}_{s+--+}}&=&{\one}\,{}\Biggl ({3}\,{\Lx}+{1\over 2}\,{\Ly^2}-{\Lx^2}-{2}\,{\Ly}+{4}\Biggr ){}+{1\over 2}\,{\UUOTT }\,{\Ly^2}+{\uot }\,{\Ly}\xxx+{i\pi}\,{}\Biggl ({\UUOTT }\,{\Ly}+{\uot }+{\one}\,{}\Biggl ({}-{2}\,{\Lx}+{\Ly}+{1}\Biggr ){}\Biggr ){}\\
{C^{(1),[1]}_{s+--+}}&=&{}-{1\over 3}\,{i\pi}\,{\one}-{5\over 9}\,{\one}\\
{A^{(1),[2]}_{s+--+}}&=&{}-{1\over 2}\,{\uot }\,{\Ly}-{1\over 4}\,{\UUOTT }\,{\Ly^2}+{\one}\,{}\Biggl ({}-{13\over 18}-{1\over 4}\,{\Ly^2}+{\Ly}\Biggr ){}\xxx+{i\pi}\,{}\Biggl ({\one}\,{}\Biggl ({}-{1\over 2}\,{\Ly}-{5\over 6}\Biggr ){}-{1\over 2}\,{\UUOTT }\,{\Ly}-{1\over 2}\,{\uot }\Biggr ){}\\
{B^{(1),[2]}_{s+--+}}&=&{}-{1\over 2}\,{\uot }\,{\Ly}-{1\over 4}\,{\UUOTT }\,{\Ly^2}+{\one}\,{}\Biggl ({}-{3\over 2}\,{\Lx}+{1\over 2}\,{\Lx^2}+{\Ly}-{1\over 4}\,{\Ly^2}-{4}\Biggr ){}\xxx+{i\pi}\,{}\Biggl ({}-{1\over 2}\,{\UUOTT }\,{\Ly}-{1\over 2}\,{\uot }+{\one}\,{}\Biggl ({}-{1\over 2}+{\Lx}-{1\over 2}\,{\Ly}\Biggr ){}\Biggr ){}\\
{C^{(1),[2]}_{s+--+}}&=&{5\over 9}\,{\one}+{1\over 3}\,{i\pi}\,{\one}
\end{eqnarray}

\subsection{$q(p_2,+) + \bar Q(p_1,-)\to q(p_3,+)+\bar Q(p_4,-)$}
\begin{eqnarray}
{A^{(1),[1]}_{t+-+-}}&=&{\tou }\,{}\Biggl ({1\over 2}\,{\Lx}+{1\over 2}\,{\Lx^2}\Biggr ){}+{1\over 4}\,{\TTOUU }\,{\Lx^2}+{\one}\,{}\Biggl ({1\over 2}\,{\Lx^2}+{13\over 18}-{1\over 3}\,{\Lx}\Biggr ){}\xxx+{i\pi}\,{}\Biggl ({1\over 2}\,{\TTOUU }\,{\Lx}+{\one}\,{}\Biggl ({\Lx}+{3\over 2}\Biggr ){}+{\tou }\,{}\Biggl ({\Lx}+{1\over 2}\Biggr ){}\Biggr ){}\\
{B^{(1),[1]}_{t+-+-}}&=&{}-{1\over 2}\,{\TTOUU }\,{\Lx^2}+{\tou }\,{}\Biggl ({}-{\Lx}-{\Lx^2}\Biggr ){}+{\one}\,{}\Biggl ({}-{2}\,{\Lx}\,{\Ly}+{\Ly^2}+{\pi^2}+{4}-{3}\,{\Ly}\Biggr ){}\xxx+{i\pi}\,{}\Biggl ({}-{\TTOUU }\,{\Lx}+{\tou }\,{}\Biggl ({}-{2}\,{\Lx}-{1}\Biggr ){}+{\one}\,{}\Biggl ({}-{2}\,{\Lx}-{3}\Biggr ){}\Biggr ){}\\
{C^{(1),[1]}_{t+-+-}}&=&{\one}\,{}\Biggl ({}-{5\over 9}+{1\over 3}\,{\Lx}\Biggr ){}\\
{A^{(1),[2]}_{t+-+-}}&=&{\one}\,{}\Biggl ({}-{1\over 2}\,{\Lx^2}+{3\over 2}\,{\Ly}-{1\over 2}\,{\Ly^2}+{\Lx}\,{\Ly}-{1\over 2}\,{\pi^2}+{1\over 3}\,{\Lx}-{13\over 18}\Biggr ){}\\
{B^{(1),[2]}_{t+-+-}}&=&{1\over 4}\,{\TTOUU }\,{\Lx^2}+{\tou }\,{}\Biggl ({1\over 2}\,{\Lx}+{1\over 2}\,{\Lx^2}\Biggr ){}+{\one}\,{}\Biggl ({}-{1\over 2}\,{\Ly^2}+{\Lx}\,{\Ly}+{3\over 2}\,{\Ly}-{4}-{1\over 2}\,{\pi^2}\Biggr ){}\xxx+{i\pi}\,{}\Biggl ({1\over 2}\,{\TTOUU }\,{\Lx}+{\one}\,{}\Biggl ({\Lx}+{3\over 2}\Biggr ){}+{\tou }\,{}\Biggl ({\Lx}+{1\over 2}\Biggr ){}\Biggr ){}\\
{C^{(1),[2]}_{t+-+-}}&=&{\one}\,{}\Biggl ({}-{1\over 3}\,{\Lx}+{5\over 9}\Biggr ){}
\end{eqnarray}
\subsection{$q(p_2,+) + \bar Q(p_1,+)\to q(p_3,+)+\bar Q(p_4,+)$}
\begin{eqnarray}
{A^{(1),[1]}_{t++++}}&=&{\one}\,{}\Biggl ({1\over 2}\,{\Lx^2}+{13\over 18}-{1\over 3}\,{\Lx}\Biggr ){}\xxx+{i\pi}\,{\one}\,{}\Biggl ({\Lx}+{3\over 2}\Biggr ){}\\
{B^{(1),[1]}_{t++++}}&=&{\TTOSS }\,{}\Biggl ({}-{\Lx}\,{\Ly}+{1\over 2}\,{\Lx^2}+{1\over 2}\,{\Ly^2}+{1\over 2}\,{\pi^2}\Biggr ){}+{\tos }\,{}\Biggl
({}-{2}\,{\Lx}\,{\Ly}+{\Lx}+{\Lx^2}-{\Ly}+{\Ly^2}+{\pi^2}\Biggr ){}\xxx
+{\one}\,{}\Biggl ({}-{2}\,{\Lx}\,{\Ly}+{\Ly^2}+{\pi^2}+{4}-{3}\,{\Ly}\Biggr ){}\xxx+{i\pi}\,{\one}\,{}\Biggl ({}-{2}\,{\Lx}-{3}\Biggr ){}\\
{C^{(1),[1]}_{t++++}}&=&{\one}\,{}\Biggl ({}-{5\over 9}+{1\over 3}\,{\Lx}\Biggr ){}\\
{A^{(1),[2]}_{t++++}}&=&{\tos }\,{}\Biggl ({\Lx}\,{\Ly}-{1\over 2}\,{\Lx}-{1\over 2}\,{\Lx^2}+{1\over 2}\,{\Ly}-{1\over 2}\,{\Ly^2}-{1\over 2}\,{\pi^2}\Biggr ){}\xxx
+{\TTOSS
}\,{}\Biggl ({1\over 2}\,{\Lx}\,{\Ly}-{1\over 4}\,{\Lx^2}-{1\over 4}\,{\Ly^2}-{1\over 4}\,{\pi^2}\Biggr ){}\xxx
+{\one}\,{}\Biggl ({}-{1\over 2}\,{\Lx^2}+{3\over 2}\,{\Ly}-{1\over 2}\,{\Ly^2}+{\Lx}\,{\Ly}-{1\over 2}\,{\pi^2}+{1\over 3}\,{\Lx}-{13\over 18}\Biggr ){}\\
{B^{(1),[2]}_{t++++}}&=&{\TTOSS }\,{}\Biggl ({1\over 2}\,{\Lx}\,{\Ly}-{1\over 4}\,{\Lx^2}-{1\over 4}\,{\Ly^2}-{1\over 4}\,{\pi^2}\Biggr ){}\xxx
+{\tos }\,{}\Biggl
({\Lx}\,{\Ly}-{1\over 2}\,{\Lx}-{1\over 2}\,{\Lx^2}+{1\over 2}\,{\Ly}-{1\over 2}\,{\Ly^2}-{1\over 2}\,{\pi^2}\Biggr ){}\xxx
+{\one}\,{}\Biggl ({}-{1\over 2}\,{\Ly^2}+{\Lx}\,{\Ly}+{3\over 2}\,{\Ly}-{4}-{1\over 2}\,{\pi^2}\Biggr ){}\xxx+{i\pi}\,{\one}\,{}\Biggl ({\Lx}+{3\over 2}\Biggr ){}\\
{C^{(1),[2]}_{t++++}}&=&{\one}\,{}\Biggl ({}-{1\over 3}\,{\Lx}+{5\over 9}\Biggr ){}
\end{eqnarray}

\subsection{$q(p_2,+) +  Q(p_1,+)\to Q(p_3,+)+ q(p_4,+)$}
\begin{eqnarray}
{A^{(1),[1]}_{u++++}}&=&{\tos }\,{}\Biggl ({1\over 2}\,{\Lx}-{1\over 2}\,{\Ly}\Biggr ){}
+{\TTOSS }\,{}\Biggl ({}-{1\over 2}\,{\Lx}\,{\Ly}+{1\over 4}\,{\Lx^2}+{1\over 4}\,{\Ly^2}+{1\over 4}\,{\pi^2}\Biggr ){}\xxx
+{\one}\,{}\Biggl ({}-{1\over 2}\,{\Lx}\,{\Ly}+{1\over 4}\,{\Ly^2}-{\Lx}+{1\over 4}\,{\pi^2}+{1\over 4}\,{\Lx^2}+{13\over 18}-{5\over 6}\,{\Ly}\Biggr ){}\\
{B^{(1),[1]}_{u++++}}&=&{\TTOSS }\,{}\Biggl ({\Lx}\,{\Ly}-{1\over 2}\,{\Lx^2}-{1\over 2}\,{\Ly^2}-{1\over 2}\,{\pi^2}\Biggr ){}
+{\tos }\,{}\Biggl ({}-{\Lx}+{\Ly}\Biggr ){}\xxx
+{\one}\,{}\Biggl ({\Lx}\,{\Ly}+{1\over 2}\,{\Ly^2}+{2}\,{\Lx}-{1\over 2}\,{\pi^2}-{1\over 2}\,{\Lx^2}+{4}+{\Ly}\Biggr ){}\xxx+{i\pi}\,{\one}\,{}\Biggl ({2}\,{\Ly}+{3}\Biggr ){}\\
{C^{(1),[1]}_{u++++}}&=&{\one}\,{}\Biggl ({}-{5\over 9}+{1\over 3}\,{\Ly}\Biggr ){}\\
{A^{(1),[2]}_{u++++}}&=&{\one}\,{}\Biggl ({}-{13\over 18}-{1\over 2}\,{\Ly^2}+{1\over 3}\,{\Ly}\Biggr ){}\xxx+{i\pi}\,{\one}\,{}\Biggl ({}-{\Ly}-{3\over 2}\Biggr ){}\\
{B^{(1),[2]}_{u++++}}&=&{\TTOSS }\,{}\Biggl ({}-{1\over 2}\,{\Lx}\,{\Ly}+{1\over 4}\,{\Lx^2}+{1\over 4}\,{\Ly^2}+{1\over 4}\,{\pi^2}\Biggr ){}+{\tos }\,{}\Biggl ({1\over 2}\,{\Lx}-{1\over 2}\,{\Ly}\Biggr ){}\xxx
+{\one}\,{}\Biggl ({}-{1\over 2}\,{\Ly}-{1\over 2}\,{\Lx}\,{\Ly}-{\Lx}+{1\over 4}\,{\Lx^2}+{1\over 4}\,{\pi^2}-{4}-{1\over 4}\,{\Ly^2}\Biggr ){}\xxx+{i\pi}\,{\one}\,{}\Biggl ({}-{\Ly}-{3\over 2}\Biggr ){}\\
{C^{(1),[2]}_{u++++}}&=&{\one}\,{}\Biggl ({}-{1\over 3}\,{\Ly}+{5\over 9}\Biggr ){}
\end{eqnarray}

\subsection{$q(p_2,+) +  Q(p_1,-)\to Q(p_3,-)+ q(p_4,+)$}
\begin{eqnarray}
{A^{(1),[1]}_{u+--+}}&=&{\one}\,{}\Biggl ({}-{1\over 3}\,{\Ly}+{1\over 2}\,{\Ly^2}-{\Lx}\,{\Ly}+{1\over 2}\,{\pi^2}-{3\over 2}\,{\Lx}+{13\over 18}+{1\over 2}\,{\Lx^2}\Biggr ){}\\
{B^{(1),[1]}_{u+--+}}&=&{1\over 2}\,{\UUOTT }\,{\Ly^2}+{\uot }\,{}\Biggl ({\Ly}+{\Ly^2}\Biggr ){}+{\one}\,{}\Biggl ({2}\,{\Lx}\,{\Ly}-{\pi^2}+{3}\,{\Lx}+{4}-{\Lx^2}\Biggr ){}\xxx+{i\pi}\,{}\Biggl ({\UUOTT }\,{\Ly}+{\uot }\,{}\Biggl ({2}\,{\Ly}+{1}\Biggr ){}+{\one}\,{}\Biggl ({2}\,{\Ly}+{3}\Biggr ){}\Biggr ){}\\
{C^{(1),[1]}_{u+--+}}&=&{\one}\,{}\Biggl ({}-{5\over 9}+{1\over 3}\,{\Ly}\Biggr ){}\\
{A^{(1),[2]}_{u+--+}}&=&{\uot }\,{}\Biggl ({}-{1\over 2}\,{\Ly}-{1\over 2}\,{\Ly^2}\Biggr ){}-{1\over 4}\,{\UUOTT }\,{\Ly^2}+{\one}\,{}\Biggl ({}-{13\over 18}-{1\over 2}\,{\Ly^2}+{1\over 3}\,{\Ly}\Biggr ){}\xxx+{i\pi}\,{}\Biggl ({}-{1\over 2}\,{\UUOTT }\,{\Ly}+{\one}\,{}\Biggl ({}-{\Ly}-{3\over 2}\Biggr ){}+{\uot }\,{}\Biggl ({}-{\Ly}-{1\over 2}\Biggr ){}\Biggr ){}\\
{B^{(1),[2]}_{u+--+}}&=&{}-{1\over 4}\,{\UUOTT }\,{\Ly^2}+{\uot }\,{}\Biggl ({}-{1\over 2}\,{\Ly}-{1\over 2}\,{\Ly^2}\Biggr ){}+{\one}\,{}\Biggl ({}-{4}-{\Lx}\,{\Ly}-{3\over 2}\,{\Lx}+{1\over 2}\,{\Lx^2}+{1\over 2}\,{\pi^2}\Biggr ){}\xxx+{i\pi}\,{}\Biggl ({}-{1\over 2}\,{\UUOTT }\,{\Ly}+{\one}\,{}\Biggl ({}-{\Ly}-{3\over 2}\Biggr ){}+{\uot }\,{}\Biggl ({}-{\Ly}-{1\over 2}\Biggr ){}\Biggr ){}\\
{C^{(1),[2]}_{u+--+}}&=&{\one}\,{}\Biggl ({}-{1\over 3}\,{\Ly}+{5\over 9}\Biggr ){}
\end{eqnarray}

\subsection{$q(p_2,+) +  Q(p_1,+)\to q(p_3,+)+ Q(p_4,+)$}
The coefficients for this amplitude are obtained by exchanging $t$ and $u$ in the
coefficients for $q(p_2,+) +  Q(p_1,+)\to Q(p_3,+)+ q(p_4,+)$,
\begin{eqnarray}
{A^{(1),[1]}_{t^\prime ++++}}&=&{A^{(1),[1]}_{u ++++}} (t \leftrightarrow u)\\
{A^{(1),[2]}_{t^\prime ++++}}&=&{A^{(1),[2]}_{u ++++}} (t \leftrightarrow u) 
\end{eqnarray}

\subsection{$q(p_2,+) +  Q(p_1,-)\to q(p_3,+)+ Q(p_4,-)$}
The coefficients for this amplitude are obtained by exchanging $t$ and $u$ in the
coefficients for $q(p_2,+) +  Q(p_1,-)\to Q(p_3,-)+ q(p_4,+)$,
\begin{eqnarray}
{A^{(1),[1]}_{t^\prime +-+-}}&=&{A^{(1),[1]}_{u +--+}} (t \leftrightarrow u)\\
{A^{(1),[2]}_{t^\prime +-+-}}&=&{A^{(1),[2]}_{u +--+}} (t \leftrightarrow u) 
\end{eqnarray}
etc.

\bibliographystyle{JHEP}
\bibliography{qqQQhel}

\end{document}